\documentclass[aps,prd,twocolumn,groupedaddress,nofootinbib]{revtex4-1}

\usepackage{amsfonts}
\usepackage{psfrag}
\usepackage{graphicx}
\usepackage{grffile}
\usepackage{cancel}
\usepackage{amsmath}
\usepackage{natbib}
\usepackage{xr}
\usepackage{hyperref}
\usepackage{verbatim}
\usepackage{epstopdf}
\usepackage{subcaption}
\usepackage{multirow}
\usepackage{commath}

\usepackage{booktabs,dcolumn,caption}

\newcolumntype{d}[1]{D{.}{.}{#1}}

\RequirePackage[normalem]{ulem} 
\RequirePackage{color}\definecolor{RED}{rgb}{1,0,0}\definecolor{BLUE}{rgb}{0,0,1} 
\providecommand{\DIFaddbegin}{} 
\providecommand{\DIFaddend}{} 
\providecommand{\DIFdelbegin}{} 
\providecommand{\DIFdelend}{} 
\providecommand{\DIFaddbeginFL}{} 
\providecommand{\DIFaddendFL}{} 
\providecommand{\DIFdelbeginFL}{} 
\providecommand{\DIFdelendFL}{} 
\newcommand{\DIFscaledelfig}{0.5}
\RequirePackage{settobox} 
\RequirePackage{letltxmacro} 
\newsavebox{\DIFdelgraphicsbox} 
\newlength{\DIFdelgraphicswidth} 
\newlength{\DIFdelgraphicsheight} 
\LetLtxMacro{\DIFOincludegraphics}{\includegraphics} 
\newcommand{\DIFaddincludegraphics}[2][]{{\color{blue}\fbox{\DIFOincludegraphics[#1]{#2}}}} 
\newcommand{\DIFdelincludegraphics}[2][]{
\sbox{\DIFdelgraphicsbox}{\DIFOincludegraphics[#1]{#2}}
\settoboxwidth{\DIFdelgraphicswidth}{\DIFdelgraphicsbox} 
\settoboxtotalheight{\DIFdelgraphicsheight}{\DIFdelgraphicsbox} 
\scalebox{\DIFscaledelfig}{
\parbox[b]{\DIFdelgraphicswidth}{\usebox{\DIFdelgraphicsbox}\\[-\baselineskip] \rule{\DIFdelgraphicswidth}{0em}}\llap{\resizebox{\DIFdelgraphicswidth}{\DIFdelgraphicsheight}{
\setlength{\unitlength}{\DIFdelgraphicswidth}
\begin{picture}(1,1)
\thicklines\linethickness{2pt} 
{\color[rgb]{1,0,0}\put(0,0){\framebox(1,1){}}}
{\color[rgb]{1,0,0}\put(0,0){\line( 1,1){1}}}
{\color[rgb]{1,0,0}\put(0,1){\line(1,-1){1}}}
\end{picture}
}\hspace*{3pt}}} 
} 
\LetLtxMacro{\DIFOaddbegin}{\DIFaddbegin} 
\LetLtxMacro{\DIFOaddend}{\DIFaddend} 
\LetLtxMacro{\DIFOdelbegin}{\DIFdelbegin} 
\LetLtxMacro{\DIFOdelend}{\DIFdelend} 
\DeclareRobustCommand{\DIFaddbegin}{\DIFOaddbegin \let\includegraphics\DIFaddincludegraphics} 
\DeclareRobustCommand{\DIFaddend}{\DIFOaddend \let\includegraphics\DIFOincludegraphics} 
\DeclareRobustCommand{\DIFdelbegin}{\DIFOdelbegin \let\includegraphics\DIFdelincludegraphics} 
\DeclareRobustCommand{\DIFdelend}{\DIFOaddend \let\includegraphics\DIFOincludegraphics} 
\LetLtxMacro{\DIFOaddbeginFL}{\DIFaddbeginFL} 
\LetLtxMacro{\DIFOaddendFL}{\DIFaddendFL} 
\LetLtxMacro{\DIFOdelbeginFL}{\DIFdelbeginFL} 
\LetLtxMacro{\DIFOdelendFL}{\DIFdelendFL} 
\DeclareRobustCommand{\DIFaddbeginFL}{\DIFOaddbeginFL \let\includegraphics\DIFaddincludegraphics} 
\DeclareRobustCommand{\DIFaddendFL}{\DIFOaddendFL \let\includegraphics\DIFOincludegraphics} 
\DeclareRobustCommand{\DIFdelbeginFL}{\DIFOdelbeginFL \let\includegraphics\DIFdelincludegraphics} 
\DeclareRobustCommand{\DIFdelendFL}{\DIFOaddendFL \let\includegraphics\DIFOincludegraphics} 

\begin{document}

\graphicspath{ {images/} }

\title{Modifying PyUltraLight to model scalar dark matter with self-interactions}

\author{Noah Glennon}
\email{nglennon@wildcats.unh.edu}
\author{Chanda Prescod-Weinstein}
\email{chanda.prescod-weinstein@unh.edu}
\affiliation{Department of Physics and Astronomy, University of New Hampshire, Durham, New Hampshire 03824, USA}

\begin{abstract}
We introduce a modification of the \textsc{PyUltraLight} code that models the dynamical evolution of ultralight axionlike scalar dark matter fields. Our modified code,  \textsc{PySiUltraLight}, adds a quartic, self-interaction term to reflect the one which arises naturally in axionlike particle models. Using a particle mass of $10^{-22}~\mathrm{eV}/\mathrm{c}^2$, we show that \textsc{PySiUltraLight} produces spatially oscillating solitons, exploding solitons, and collapsing solitons which prior analytic work shows will occur with attractive self-interactions. Using our code we calculate the oscillation frequency as a function of soliton mass and equilibrium radius in the presence of attractive self-interactions. We show that when the soliton mass is below the critical mass ($M_c = \frac{\sqrt{3}}{2}M_{\mathrm{max}}$) described by Chavanis [Phys. Rev. D \textbf{94}, 083007 (2016)] and the initial radius is within a specific range, solitons are unstable and explode.  We test the maximum mass criteria described by Chavanis [Phys. Rev. D \textbf{94}, 083007 (2016)] and Chavanis and Delfini [Phys. Rev. D \textbf{84}, 043532 (2011)] for a soliton to collapse when attractive self-interactions are included.  We also analyze both binary soliton collisions and a soliton rotating around a central mass with attractive and repulsive self-interactions.  We find that when attractive self-interactions are included, the density profiles get distorted after a binary collision. We also find that a soliton is less susceptible to tidal stripping when attractive self-interactions are included. We find that the opposite is true for repulsive self-interactions in that solitons would be more easily tidally stripped.  Including self-interactions might therefore influence the survival timescales of infalling solitons.
\end{abstract}

\maketitle

\section{Introduction}

There is significant evidence to show that dark matter makes up the majority of the matter in the Universe and that this evidence suggests that dark matter is nonrelativistic and collisionless ~\cite{Buckley2018,Ade2015,Schwabe2016,Armendariz2014}. Strong evidence for this cold dark matter (CDM) picture comes from fluctuations in the cosmic microwave background~\cite{Giesen2012,Aghanim2020}. We still know very little about the microphysics of the particle or particles that comprise dark matter.  One of the most successful and well-studied dark matter paradigms is weakly interacting massive particles (WIMPs). WIMPs are consistent with the statistics of large-scale structures formed in the Universe.  However, at small scales, questions remain. Models, such as the WIMP model and other cold, collisionless models, with only dark matter predict dense cusps at the center of dark matter halos.  Observations suggest that many dark matter halos have constant density cores.  This apparent issue may be resolved with the inclusion of baryonic processes such as supernova feedback, tidal stripping, and dynamical friction~\cite{Popolo2016,Marsh2015}. 

Traditional CDM models like WIMPs faced the missing satellite problem, which is the discrepancy between the few subhalos we observed and the number found in CDM simulations.  There are several proposed theoretical solutions to the problem~\cite{Buckley2018,Spergel2000,Avila_Reese2001,Kamionkowski2000}.  As astronomers find more satellite galaxies, there is increasing confidence that the missing satellite problem is no longer a real problem. WIMPs also face the ``too big to fail" problem, which comes from the most dense dwarf galaxy-hosting subhalos predicted by simulations being systematically denser than what we infer from the brightest Milky Way satellite galaxies~\cite{Papastergis2014}. This problem might also be solved by including baryonic effects such as supernova feedback and tidal stripping~\cite{Popolo2016,Marsh2015}. 

Another possible solution is that an alternative to the CDM paradigm is needed.  Axionlike particles (ALPs), which are motivated by open problems in QCD and also can come from string theory compactifications, are one such dark matter candidate~\cite{Schwabe2016,Buckley2018,Arvanitaki2010,Cicoli2013}. This class of scalar dark matter particles has its roots in the QCD axion, which is a consequence of the Peccei-Quinn mechanism that was developed to solve the \textit{CP} problem in QCD. The axion arises in this model through spontaneous symmetry breaking as a psuedo-Nambu-Goldstone boson~\cite{Peccei1977}.  The QCD axion potential is given by
\begin{equation}
    V(\phi) = \Lambda^4 \left( 1-\mathrm{cos}(\phi/f_a) \right),
\end{equation}
where $f_a$ is the Peccei-Quinn symmetry breaking scale and $\Lambda \approx 0.1~\mathrm{GeV}$~\cite{Guth2015}.  Axion dark matter models can give the expected dark matter abundance we see in the Universe~\cite{Dine1983,Preskill1983,Abbott1983,Kim2010}.  From~\cite{Guth2015}, the abundance of QCD axion dark matter is given by:
\begin{equation}
    \Omega_a \approx \left( \frac{f_a}{10^{11-12}~\mathrm{G e V}} \right)^{7/6}.
\end{equation}

In the literature, the QCD axion is typically theorized to be more massive than the ALPs that are motivated by string compactifications because the QCD $f_a$ cannot be trans-Planckian.  From~\cite{DiLuzio2020} the QCD axion mass is given by:
\begin{equation}
    m_a \simeq 5.7\left(\frac{10^{12}~\mathrm{GeV}}{f_a}\right)\mu\mathrm{eV}.
\end{equation}
It is important to note that this relation only applies to the QCD axion and not necessarily to ALPs.  The mass of the QCD axion has a range of values but is estimated to be between $10 - 10^3~\mu \mathrm{eV}$~\cite{Alesini2019}.  The lower limit is set because the QCD $f_a$ cannot be trans-Planckian.  This is because the shift symmetry for the axion would be unbroken if the decay constant were above the Planck scale~\cite{Rudelius2015}.

In this paper, we focus on the physics of ultralight axions and not the QCD axion.  ALPs are also sometimes referred to as ultralight axions (ULAs). Fuzzy dark matter is a ULA model where dark matter comprises ultralight bosons $\sim \mathcal{O}(10^{-22})~\mathrm{eV}/\mathrm{c}^2$~\cite{Hu2000, Hui2017,Ringwald2012}. Because the mass of fuzzy dark matter is very small, the de Broglie wavelength is large (on the order of a kiloparsec) which means that dense cusps would be suppressed by quantum pressure.  Here, quantum pressure is a term used to describe the effective particle-particle interaction that is found in the Madelung formalism~\cite{Madelung1926}.  Due to the wavelike nature of ULA models, on small scales (on the order of a few kpc), we should observe interference patterns~\cite{Schive2014,Schive2014_2,Schwabe2016}.  ULA models have similar predictions to WIMP models on large scales. However, on small scales, ULA models have suppressed structure formation~\cite{Peebles2000}.  In particular, ULA models will not have dark matter halos on scales smaller than the Jeans scale~\cite{Hu2000,Suarez2018}. Focusing on small-scale structures is therefore important to test the ULA hypothesis. ULAs leads to distinct phenomenological considerations from those with more massive ALPs, which can lead to miniclusters of cosmological interest, see e.g.,~\cite{2019JCAP...04..012V,Kavanagh2020}. 
 
 ULA models that include self-interactions may diverge from those that are gravity only~\cite{Chavanis2016}.  If the self-interaction is attractive, as is the case with the QCD axion, virialized dark matter clusters of sufficient mass may collapse into black holes, explode, or oscillate in size.  Supermassive black holes could be formed by ultralight particles such as ALPs with an attractive self-interaction.   Meanwhile, the presence of a repulsive self-interaction guarantees the existence of a stable halo configuration in the Newtonian limit~\cite{Chavanis2016}. This configuration occurs when the repulsive self-interaction and repulsive quantum pressure balance gravity's attractive behavior.  In the scenario where the scalar dark matter has a repulsive interaction,~\cite{Chavanis2016,Fan2016} suggest that observations of the bullet cluster sets a limit on the repulsive self-interaction strength.  This constraint combined with the mass of one of the lightest known dark matter halos, Willman I, we can estimate the mass of the boson to be on the order of $m = 10^{-2}~\mathrm{eV}/\mathrm{c}^2$ which is too massive to fall into the category of ultralight dark matter~\cite{Chavanis2016}.

To better quantitatively understand the effects of self-interactions on ULA models, we introduce \textsc{PySiUltraLight}, a modified version of \textsc{PyUltraLight} which simulates the dynamics of axion fields~\cite{Edwards2018}, which solves the Schrödinger-Poisson equations.  Our code upgrades the original by including the self-interaction term which accounts for additional physics.  We validate the code by verifying the analytic predictions made in~\cite{Chavanis2016}. In particular, we confirm the oscillation behavior of solitons, the conditions under which exploding solitons will occur, and the existence of a maximum soliton mass given attractive self-interactions.  Our code is also capable of including repulsive self-interactions.  We explore the behavior of solitons in binary collisions and in orbits around a central potential when there are attractive or repulsive  self-interactions.  

The results of this paper are of interest because we show the phenomenological differences between ultralight dark matter models with and without self-interactions.  Understanding these differences allows for more detailed simulations to better understand how ultralight dark matter models contrast with CDM models.  Since there is no reason dark matter cannot have a self-coupling, it is unlikely the self-interaction strength is exactly zero.  Most well-motivated particle physics models include a self-interaction.  Since the coupling is typically small, astrophysicists tend to neglect the coupling in simulations.  However, in this paper, we show that including the self-interactions affects the phenomenology of solitary and binary soliton systems, including mergers.  Self-interactions are believed to affect the density-radius relationship of solitonic dark matter cores~\cite{Eby2020}.  Understanding how solitons oscillate may also be important for understanding dark matter structure formation because for a range of axion masses, these objects may decay in the matter-radiation equality or matter dominated epochs~\cite{Olle2019,Kawasaki2019}.

The contents of the paper are as follows. Section~\ref{fdmeom} outlines the physics of ULA models with the inclusion of self-interactions and describes the Gross-Pitaevskii-Poisson (GPP) equations in dimensional and adimensional forms.  Section~\ref{implementation} shows how the equations of motion are implemented in \textsc{PySiUltraLight}.  In Sec.~\ref{solitonic}, we discuss results found with \textsc{PySiUltraLight}.  Using a Gaussian ansatz, we find the oscillation frequencies of a soliton with attractive self-interactions and test the conditions under which exploding solitons can occur.  We also include an analysis for finding the maximum mass for a soliton with attractive self-interactions.  In Sec.~\ref{msolitonic}, we discuss simulations of both binary soliton collisions and solitons rotating around a central potential with both repulsive and attractive self-interactions. Our conclusion that the self-interaction remains an important consideration---and future directions---are discussed in Sec.~\ref{conclusions}.


\section{ULA Equations of Motion}
\label{fdmeom}
ULA dark matter in the high occupancy regime---e.g., Bose-Einstein condensate-like solitons---are normally described by the Schrödinger-Poisson equations~\cite{Guth2015,Marsh2016,Baldeschi1983,Sin1994,Lee1996,Matos2000,Peebles2000,Lee2015,Arbey2001,Alcubierre2001,Silverman2002,Arbey2003,Bohmer2007,Fukuyama2008,Sikivie2009,Lee2010,Ruffini1969}. When self-interactions are included, the Schrödinger-Poisson equations become the nonlinear Schrödinger-Poisson or Gross-Pitaevskii-Poisson equations~\cite{Chavanis2016,Chavanis2011}. We use the term soliton to refer to self-localized axion dark matter which is near its ground state in the Newtonian limit~\cite{Amin2019,Ruffini1969}.  We focus on the leading order self-interaction term which is a quartic term and do not consider higher order terms. Assuming a classical field with minimal coupling to gravity, the action takes the form:
\begin{equation}
    S = \int d^4 x \sqrt{-g} \left[\frac{1}{2} g^{\mu\nu} \partial_{\mu}\phi\partial_{\nu}\phi-\frac{1}{2}m^2\phi^2-\frac{\lambda}{4}\phi^4\right]. 
\end{equation}
Here, $\phi$ is the scalar field, $m$ is the mass of the scalar field, and $\lambda$ is the dimensionless self-coupling strength.  

The self-coupling strength, or self-interaction strength, is also directly proportional to the scattering length, $a_s$.  The scattering length is a measure of the interaction cross section.  When $a_s$ is positive, the interaction is repulsive.  When $a_s$ is negative, the interaction is attractive~\cite{Bao2013}. $\lambda$ may be written in terms of the scattering length $a_s$.  In~\cite{Chavanis2016},  $\lambda$ is given by
\begin{equation}
    \lambda = \frac{8\pi a_s m c}{\hbar} = \frac{8 \pi M_p^2 G a_s m}{\hbar^2}
\end{equation}
where $M_p$ is the Planck mass and $m$ is the scalar field mass. $\lambda$ can also be defined in terms of the decay constant, $f_a$, with the relation $\lambda = \frac{m^2}{f_a^2}$~\cite{Desjacques2017}. Constraints from observations require a very small ULA self-interaction strength. According to~\cite{Fan2016}, constraints from the bullet cluster require that the repulsive coupling be
\begin{equation}
    \lambda < 10^{-11} \left( \frac{m}{\mathrm{eV}/\mathrm{c}^2} \right)^{3/2}.
\end{equation}

We find the equations of motion in the Newtonian gauge, and writing the real scalar field $\phi$ in terms of a complex field $\psi$ by
\begin{equation}
    \phi = \frac{\hbar}{\sqrt{2m}}\left(\psi e^{-i m t/\hbar} + \psi^* e^{i m t/\hbar}\right)
\end{equation}
(see~\cite{Kirkpatrick2020}), we arrive at the GPP equations:

\begin{equation}
	i \hbar \dot{\psi} = -\frac{\hbar^2}{2m} \nabla^2 \psi + m \Phi \psi + \frac{4\pi \hbar^2 a_s}{m} \abs{\psi}^2 \psi
	\label{eq:GPeqn}
\end{equation}
and
\begin{equation}
	\nabla^2 \Phi = 4\pi G m \abs{\psi}^2.
	\label{eq:Peqn}
\end{equation}
Here, $\psi$ is the boson field, and $\Phi$ is the gravitational potential.  Using the process found in~\cite{Edwards2018}, the nondimensional form of the GPP equations is
\begin{equation}
	i \dot{\psi} = -\frac{1}{2} \nabla^2 \psi + \Phi \psi + \kappa \abs{\psi}^2 \psi
	\label{eq:aGPeqn}
\end{equation}
and
\begin{equation}
	\nabla^2 \Phi = 4\pi \abs{\psi}^2.
	\label{eq:aPeqn}
\end{equation}

Here we have introduced  $\kappa$, a dimensionless coupling constant that will simplify our numerical efforts later.  By definition,
\begin{equation}
	\kappa = \frac{4\pi \hbar a_s}{\mathcal{T}m^2 G},
	\label{eq:kappa}
\end{equation}
where $\mathcal{T}$ is the timescale given in~\cite{Edwards2018} as
\begin{equation}
    	\mathcal{T} =  \left(\frac{8\pi}{3H_0^2\Omega_{m0}}\right)^{\frac{1}{2}} \approx 75.5\mathrm{Gyr}.
\end{equation}

For the equilibrium solution for a system of self-gravitating bosons, we find the solution in the Newtonian limit.  A general relativistic treatment can be found in~\cite{Ruffini1969}.  It is important to note some differences in the general relativistic treatment.  For instance, in the general relativistic treatment, there is always a maximum soliton mass, even when there are repulsive self-interactions~\cite{Chavanis2016}.  Validating that the Newtonian limit is a good assumption is important for our simulations.  Appendix E in~\cite{Chavanis2016} shows that the Newtonian limit is valid for our regime.  They show that the Newtonian limit is a good approximation except when a collapsing soliton is very close to the end of its collapse time.

 The significance of including the self-interaction merits discussion, since as we described earlier, typically ULA models, and more generally scalar dark matter models, ignore self-interactions between the particles in favor of considering only their coupling through gravity.  This is because the dimensionless coupling is on the order of $10^{-96}$~\cite{Desjacques2017}.  However, even though the coupling is very small, to understand effective coupling strength, we must multiply this value by the phase space density of axions in the environment~\cite{Desjacques2017}. One way to have some intuition for this is to consider that the density associated with the $\psi$ field is given by $\rho = m\abs{\psi}^2$. This means the nonlinear term $\kappa \abs{\psi}^2\psi \propto \kappa \rho \psi$. In other words, the effect of the nonlinear term is governed by the combination of $\kappa$ and the density.
 
 This is not a surprising conclusion since this result arises in other areas of physics. For example, in optical systems that are governed by nonlinear Schrödinger/Gross-Pitaevskii equation, the nonlinear term is responsible for inducing an intensity-dependent refractive index. This change in the refractive index is also known as the Kerr effect ~\cite{Agrawal2019}. Therefore, there is good reason to take seriously the presence of a self-interaction, even if it is small. Results in the following sections are consistent with this expectation.


\section{Implementation in PySiUltraLight}
\label{implementation}

\textsc{PyUltraLight} is a Python solver for the evolution of solitonic dark matter and solves the Schrödinger-Poisson system of equations.  The code uses a pseudospectral solver where the linear differential operators are computed in Fourier space whereas the nonlinear terms are computed in phase space.  We made the following changes to the code to adapt it for a self-coupling term, creating a version we refer to as \textsc{PySiUltraLight}.

The original \textsc{PyUltraLight} paper \cite{Edwards2018} goes over how to implement the Schrödinger-Poisson equations as well as how to make a soliton profile used for simulations.  \textsc{PyUltraLight} uses a soliton profile of self-gravitating bosons in equilibrium.  The method used to get the profile can be found in more detail in~\cite{Edwards2018} or~\cite{Ruffini1969}. Reference \cite{Edwards2018} then discusses several simulations of binary soliton collisions and solitons rotating around a central potential.  We produce similar but distinct simulations which now include self-interactions. The results appear in Sec.~\ref{msolitonic}.  Reference \cite{Edwards2018} also details how convergence tests were performed to verify energy conservation and to see how spatial and temporal resolutions affect the simulation results.

To find the dynamics of solitons, we follow a similar procedure as in~\cite{Edwards2018}. We rederived the evolution equations in the presence of self-interactions.  First, we determined how $\psi$ changes when a small step in time is taken:

\begin{multline}
	\psi (\vec{x},t+h) = T \exp \left[-i \int^{t+h}_t dt' \left( -\frac{1}{2} \nabla^2 + \Phi(\vec{x},t')\right.\right.\\
	+ \left.\left.\kappa \abs{\psi(\vec{x},t')}^2 \right)\right] \psi(\vec{x},t).
	\label{eq:timestep}
\end{multline}
Here, $T$ is the time ordering operator.  Next we use the following approximations for small time steps which come from the trapezoid rule

\begin{equation}
	\int^{t+h}_t dt' \Phi(\vec{x},t') \approx \frac{h}{2}\left( \Phi(\vec{x},t+h) + \Phi(\vec{x},t)\right)
\end{equation}
	
\begin{equation}
	\int^{t+h}_t dt' \kappa \abs{\psi(\vec{x},t')}^2 \approx \frac{\kappa h}{2}\left( \abs{\psi(\vec{x},t+h)}^2 + \abs{\psi(\vec{x},t)}^2\right).
\end{equation}
Substituting these into Eq.~(\ref{eq:timestep}) gives

\begin{multline}
	\psi(\vec{x},t+h) \approx \exp\left[-\frac{ih}{2} \Phi(\vec{x},t+h)\right] \\\exp\left[-\frac{ih\kappa}{2}\abs{\psi(\vec{x},t+h)}^2\right]\\
\exp\left[-\frac{ih}{2}\nabla^2\right] \exp\left[-\frac{ih}{2} \Phi(\vec{x},t)\right] \\\exp\left[-\frac{ih\kappa}{2}\abs{\psi(\vec{x},t)}^2\right] \psi(\vec{x},t).
\end{multline}

Using the Baker-Campbell-Hausdorf formula we verified that the error is $\mathcal{O}(h^3)$.  This is the same order of error as when there were no self-interactions (i.e. when $\kappa = 0$).  The procedure to evolve the soliton uses the following equations:

\begin{multline}
	\psi(\vec{x},t+h) = \exp\left[-\frac{ih}{2} \Phi(\vec{x},t+h)\right] \\\exp\left[-\frac{ih\kappa}{2}\abs{\psi(\vec{x},t+h)}^2\right] \\
	\mathcal{F}^{-1}\exp\left[-\frac{ih}{2}k^2\right]
	\mathcal{F} \exp\left[-\frac{ih}{2} \Phi(\vec{x},t)\right]\\
	 \exp\left[-\frac{ih\kappa}{2}\abs{\psi(\vec{x},t)}^2\right] \psi(\vec{x},t)
\end{multline}

\begin{equation}
	\Phi(\vec{x},t+h) = \mathcal{F}^{-1}\left(-\frac{1}{k^2}\right)  \mathcal{F}4\pi \abs{\psi(\vec{x},t_i)}^2.
\end{equation}
Here, $\mathcal{F}$ is the Fourier transform, $ \mathcal{F}^{-1}$ is the inverse Fourier transform, and $k$ is the wave number in Fourier space.  $\psi(\vec{x},t_i)$ is the field at the half step. We checked energy conservation to verify the integrity of changes we made to the code.  The Lagrangian density that yielded the GPP equations is
\begin{multline}
	\mathcal{L} = -\left(\frac{1}{2}\abs{\nabla \Phi}^2 + \Phi \abs{\psi}^2 + \frac{1}{2} \abs{\nabla \psi}^2 +\right.\\
	\left.\frac{i}{2}\left(\psi \dot{\psi^*} - \dot{\psi}\psi^*\right) + \frac{\kappa}{2}\abs{\psi}^4\right).
\end{multline}
The energy of the system then becomes

\begin{equation}
	E_{tot} = \int_V d^3x \left(\frac{1}{2} \Phi \abs{\psi}^2 - \frac{1}{2} \psi^* \nabla^2 \psi + \frac{\kappa}{2}\abs{\psi}^4\right).
\end{equation}
\textsc{PySiUltraLight} follows \textsc{PyUltraLight} in using the following units. We will often refer to these units, which take the following form, as code units: 

\begin{equation}
	\mathcal{L} = \left(\frac{8\pi \hbar^2}{3m^2 H_0^2 \Omega_{m0}}\right)^{\frac{1}{4}} \approx 121 \left(\frac{10^{-23}~\mathrm{eV}}{m}\right)^{\frac{1}{2}} \mathrm{kpc},
\end{equation}

\begin{equation}
    	\mathcal{T} =  \left(\frac{8\pi}{3H_0^2\Omega_{m0}}\right)^{\frac{1}{2}} \approx 75.5~\mathrm{Gyr},
\end{equation}

and

\begin{multline}
  	  \mathcal{M} = \frac{1}{G}\left(\frac{8\pi}{3H_0^2\Omega_{m0}}\right)^{-\frac{1}{4}}\left(\frac{\hbar}{m}\right)^{\frac{3}{2}} \\
  	  \approx 7\times10^7 \left(\frac{10^{-23}~ \mathrm{eV}}{m}\right)^{\frac{3}{2}}M_\odot.
\end{multline}

These are the code length, time, and mass respectively where $H_0$ is the present-day Hubble parameter and $\Omega_{m0}$ is the present-day matter fraction of the energy density of the Universe. One can get dimensionful quantities back by using dimensional analysis.  One can recover the desired unit by taking a code unit and multiplying it by the proper code quantities.  For example, to recover the dimensional form of the energy

\begin{equation}
    E = \mathcal{M}\mathcal{L}^2\mathcal{T}^{-2}E_{code}.
\end{equation}

To initialize \textsc{PySiUltraLight}, users must specify the initial soliton profiles by giving the solitons the desired masses, positions, velocities, and phases.  \textsc{PySiUltraLight} additionally requires that the user choose the boson mass and the dimensionless coupling constant which are necessary parameters for determining the solitonic dynamics with self-interactions.  Other parameters that must be specified that are also necessary in \textsc{PyUltraLight} are the box size (which is the size of the simulation), the duration of the simulation, and the time step.  The time step used in the code is independent of the self-interaction strength.  It is only dependent on the resolution but may also be adjusted with the step factor.

\textsc{PySiUltraLight} uses periodic boundary conditions from \textsc{PyUltraLight} which can, in principle, affect this code's results. Using periodic boundary conditions means that the space in which the simulation runs is topologically equivalent to a torus. When mass travels through the boundary, unphysical affects can arise because angular momentum is not conserved. There may also be unwanted effects when the box size is too small relative to the radius of the soliton. When this happens, the soliton may interact with its neighboring images in an unphysical way.  In all of our simulations, the solitonic bodies never cross the boundary.  For the simulations presented in this paper, we also checked to see if increasing the box size made any changes to the results.  In all cases, the differences were negligible.


\section{Single Soliton Behavior in PySiUltraLight}
\label{solitonic}

There are noticeable differences between simulations with large attractive self-interactions and those without self-interactions, indicating that it can be physically important to account for the presence of self-interactions.  We focus on simulations with attractive self-interactions in this section to test the analysis in~\cite{Chavanis2016}. For portions of this work, we will employ a Gaussian ansatz for the soliton profile. The ansatz approximates the exact profile and is commonly used in studying Bose-Einstein-condensates (BECs)~\cite{Chavanis2016, Chavanis2011, Chavanis2011_2}, and it is an approximate solution to the Gross-Pitaevskii-Poisson equations~\cite{Chavanis2016}. We are motivated to use the Gaussian ansatz in order to make direct comparisons with~\cite{Chavanis2016}. This approximation involves defining a density profile for a self-gravitating BEC and then minimizing the total energy of the system with respect to the radius to find the equilibrium radius. This technique has also been used to describe the mass-radius relation in white dwarf stars~\cite{Chavanis2011_2}. The form for the soliton using the Gaussian ansatz is
\begin{equation}
	\psi(r,t) = \left[\frac{M}{\pi^{3/2}R(t)^3}\right]^{1/2}e^{-\frac{r^2}{2R(t)^2}}e^{imH(t)r^2/2\hbar}
\end{equation}
where $R(t)$ is the measure of the size of the soliton, $r$ is the distance from the center of the soliton, and $H(t) = \frac{\dot{R}}{R}$.  This should not be confused with the Hubble parameter.  The dependence on the self-interaction is included in $H(t)$. There are other profiles we could have used~\cite{Eby2018}, however, we wanted to compare our results with the claims in~\cite{Chavanis2016} which used a Gaussian ansatz.  This profile is evolved using the GPP equations using the methods explained in Sec.~\ref{implementation}.

\textsc{PyUltraLight} makes a soliton profile by imposing spherical symmetry on the Schrödinger-Poisson equations and also requires that the radial density profile be time independent~\cite{Edwards2018}. This is thought to be a good approximation for undisrupted solitonic cores (see e.g.~\cite{Edwards2018,Ruffini1969}). We use the ansatz when making comparisons to the analytical work found in~\cite{Chavanis2016}.  Specifically, we use the ansatz when looking at oscillating and exploding solitons.  \textsc{PySiUltraLight} expands beyond its predecessor code by adding the ability for the user to specify the radius of the soliton when there are attractive self-interactions. This is necessary to implement the Gaussian ansatz and more broadly introduces the capability to change the radius of a soliton. 

The first step to verify the newly adjusted program was working properly involved comparing \textsc{PySiUltraLight} with zero self-coupling to \textsc{PyUltraLight}. There were no differences between the new and old versions in the tests that we ran with this condition.  The next step in verifying the code was to ensure the energy of the system was conserved. Energy was indeed conserved to a high degree, provided that the time step was sufficiently small.  The degree to which energy was conserved also depended on the resolution of the simulation.

In this section, we will first show the oscillatory behaviors of single solitons.  We will then describe the behaviors of exploding solitons and the conditions in which they should be observed.  Lastly, we present numerical work on collapsing solitons, which requires going beyond the Gaussian ansatz and results available in earlier literature, e.g., ~\cite{Chavanis2016}.

\subsection{Oscillating solitons}

We considered~\cite{Chavanis2016} as a benchmark for testing, and our next step was to show that \textsc{PySiUltraLight} outputs results which match those found there.  In~\cite{Chavanis2016,Eby2016,Chen2020}, axion dark matter with an attractive self-coupling has a maximum halo mass before the dark matter collapses into a black hole

\begin{equation}
	M_{\mathrm{max}} = 1.012 \frac{\hbar}{\sqrt{G m \abs{a_s}}},
	\label{eq:Mmax}
\end{equation}
which can be rewritten as

\begin{equation}
	\frac{M_{\mathrm{max}}}{M_\odot} = 1.56\times 10^{-34} \left(\frac{\mathrm{eV}/\mathrm{c}^2}{m}\right)^{\frac{1}{2}} \left(\frac{\mathrm{fm}}{\abs{a_s}}\right)^{\frac{1}{2}}.
	\label{eq:maxmass}
\end{equation}
This means there are no time-independent solutions to the GPP equations when the soliton mass is too large and the self-interaction is attractive.  

Using the Gaussian ansatz, we can test how solitons will oscillate given a specified mass and radius.  There are two different scenarios where one can see oscillating solitons.  The first scenario is when the soliton mass is below a critical mass $M_c = \frac{\sqrt{3}}{2}M_{\mathrm{max}}$.  When this is the case, the soliton oscillates provided the starting radius is larger than
\begin{equation}
    R = \frac{1+\sqrt{1-(\frac{M}{M_c})^2}}{2M}.
\end{equation}
In the other scenario, the soliton mass is between the critical mass and the maximum mass.  In this case, there is a finite range that the initial radius can be, otherwise the soliton will collapse. For oscillating solitons, there is a relation between the oscillation frequency and the equilibrium radius when oscillations are small.  According to~\cite{Chavanis2016}, the relation is given by
\begin{equation}
	\omega^2 = \frac{2\left(R^2_e - 1\right)}{R^4_e \left(R^2_e + 1\right)}.
	\label{eq:radrelation}	
\end{equation}
Here, $R_e$ is the equilibrium radius in terms of the radius of the soliton with a mass of $M_{\mathrm{max}}$ and the frequency $\omega$ had been scaled by the dynamical time.  The equilibrium radius is the radius where the effective potential as a function of radius is minimized.  The equilibrium radius is given by
\begin{equation}
	R_e = \frac{1 \pm \sqrt{1-M^2}}{M},	
\end{equation}
where $M$ is the total soliton mass~\cite{Chavanis2016}.  The dynamical time is given by
\begin{equation}
	t_D = 3\sqrt2 \frac{\abs{a_s}\hbar}{G m^2}.	
\end{equation}

\begin{figure}
	
	\centering
	{\includegraphics[width=.5\textwidth]{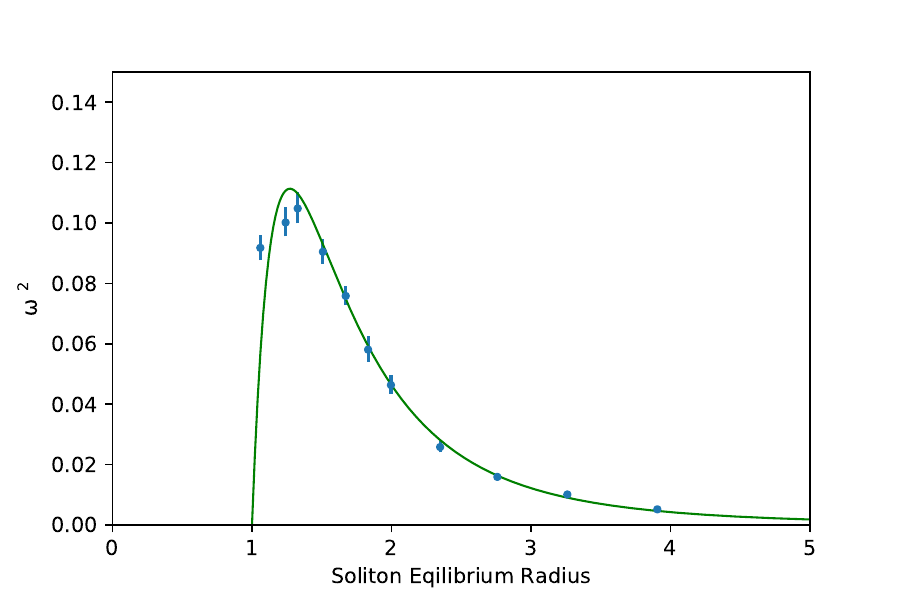}}
	\caption{This plot shows how changing the equilibrium radius changes the frequency of oscillation.  The green line shows the fit of Eq.~(\ref{eq:radrelation}) scaled by the best fit dynamical time squared.}
	\label{fig:oscradfreq}
\end{figure}

We ran several simulations to determine the frequency of oscillations by analyzing the time for the soliton to go from a peak density to the next peak density.  We do this because the density peak will occur when the soliton is smallest in size.  For these simulations, the only parameter varied was the total soliton mass between $0.3M_{\mathrm{max}} < M < M_{\mathrm{max}}$ so as to isolate the relationship between the mass and the frequency of oscillation.  The boson mass was $m = 1 \times 10^{-22}~ \mathrm{eV}/\mathrm{c}^2$, and the coupling was $\kappa = -2.0$.  Figure~\ref{fig:oscradfreq} shows the plot of the soliton equilibrium radius versus the oscillation frequency.  The line in the plot shows the best fit of Eq.~(\ref{eq:radrelation}) which fits the data well.  The data does not match the best fit line as well when the mass approaches the maximum soliton mass.

\begin{figure}
	
	\centering
	{\includegraphics[width=.5\textwidth]{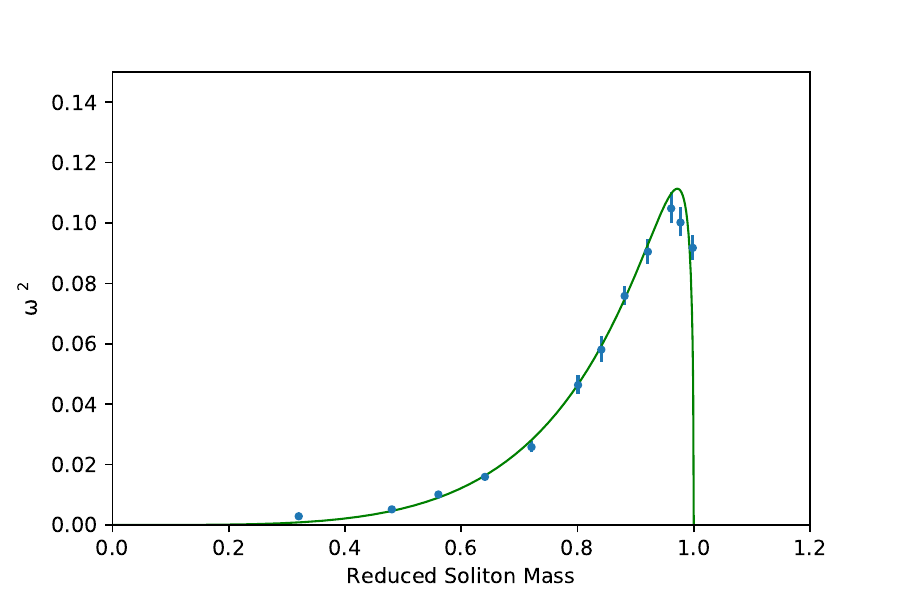}}
	\caption{This plot shows changing the equilibrium radius changes the frequency of oscillation.  The green line shows the fit of Eq.~(\ref{eq:massrelation}) scaled by the best fit dynamical time squared.}
	\label{fig:oscmassfreq}
\end{figure}

The soliton oscillation frequency can also be found using total mass.  Figure~\ref{fig:oscmassfreq} shows the soliton mass versus the oscillation frequency.  For positive $\omega^2$, the equation is given by
\begin{equation}
	\omega^2 = \frac{2M^4\left(\sqrt{1-M^2}-M^2+1\right)}{\left(\sqrt{1-M^2}+1\right)^5}	
	\label{eq:massrelation}.
\end{equation}

\begin{figure}

	\centering
	{\includegraphics[width=.45\textwidth]{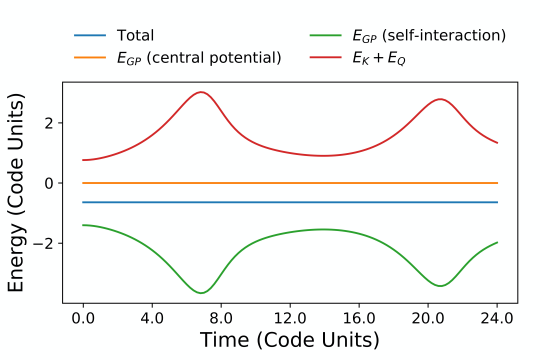}}
	\caption{The energy components (the total energy, the energy associated with a central potential, the gravitational potential energy, and the kinetic and quantum energies) of an oscillating soliton system. Note that the total energy is conserved (to within 1 part in $10^4$).}
	\label{fig:oscillateenergy}
\end{figure}

\begin{figure}
\begin{tabular}{cc}
\includegraphics[width=.31\textwidth]{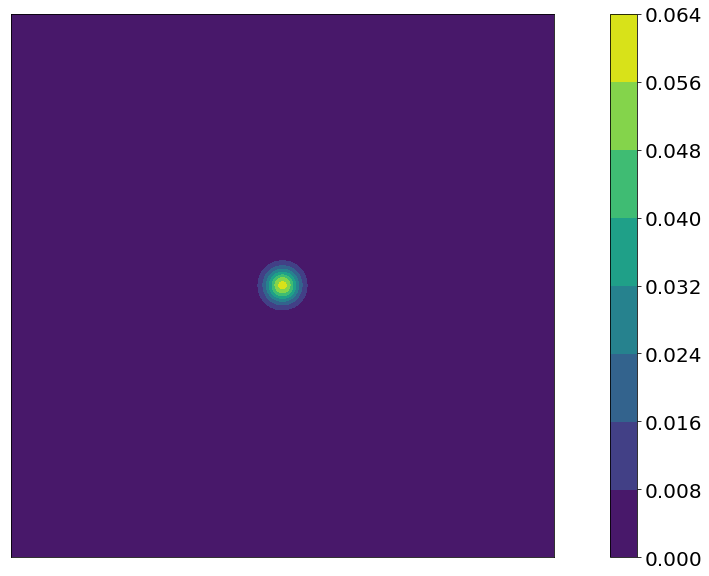}\\
(a) t = 1.5 \\
\includegraphics[width=.31\textwidth]{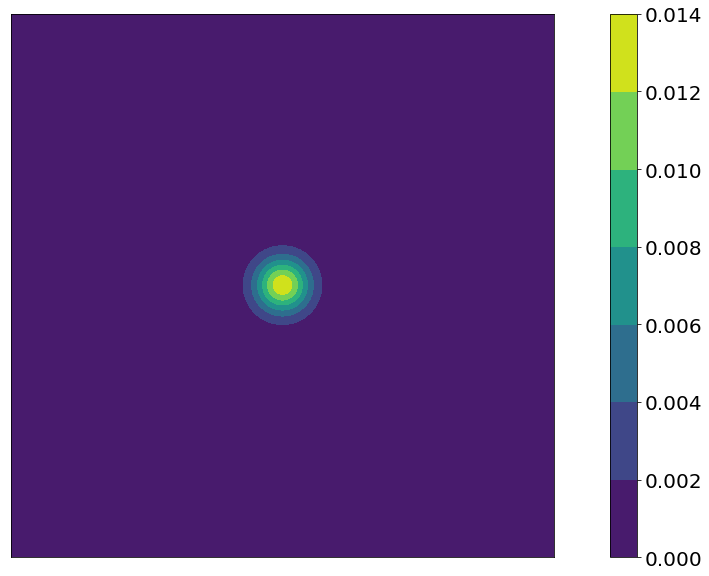}\\
(b) t = 3.0 \\ 
\includegraphics[width=.31\textwidth]{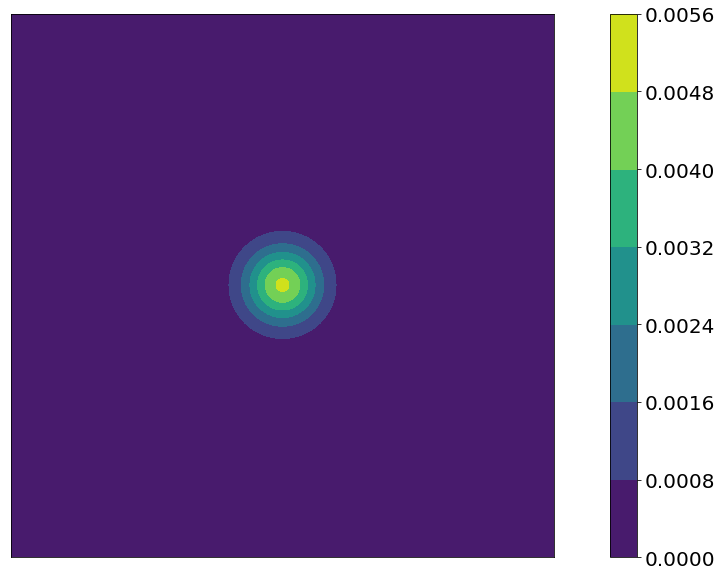}\\
(c) t = 4.5 \\
\includegraphics[width=.31\textwidth]{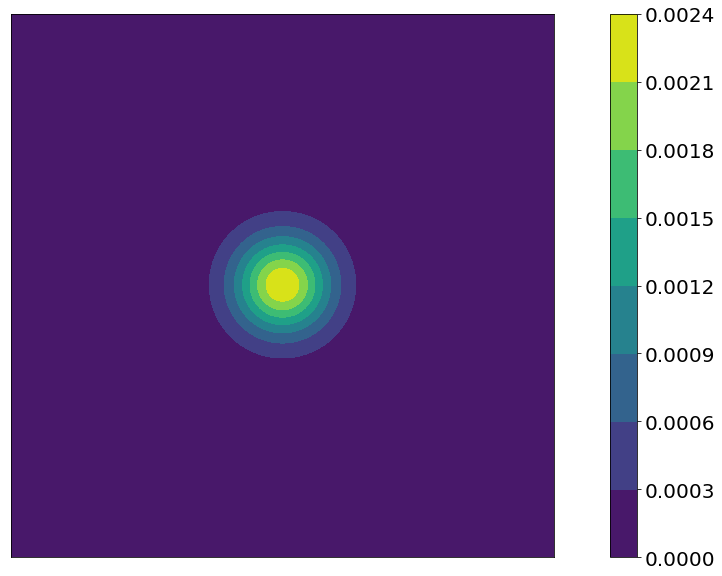}\\
(d) t = 6.0
\end{tabular}
\caption{A single soliton with a mass of 1.0 code units ($2.3\times 10^6~\mathrm{M}_\odot$) explodes. These plots show contours of constant density. Time progresses from top to bottom, and the time under each frame is indicated in code units. In the Gaussian ansatz, $R(0) = 1$ code unit.  This is well below the critical mass of about 2.3 ($5.3\times 10^6~\mathrm{M}_\odot$) code units.  Here, $\kappa = -2.0$.  The box length is 45 code units and the time step is 0.3.}
\label{fig:explode}	
\end{figure}

\begin{figure}

	\centering
	{\includegraphics[width=.45\textwidth]{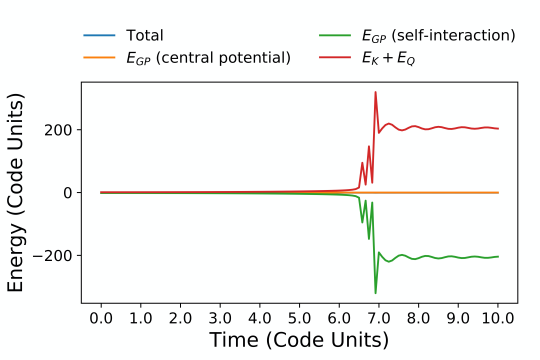}}
	\caption{Similar to Fig.~\ref{fig:oscillateenergy}: the energy components of a collapsing soliton system.  The total energy (in blue) is under the orange line.}
	\label{fig:collapseenergy}
\end{figure}

In the first simulation for testing oscillating solitons, the soliton had a mass of 2.4 code units which is less than $M_{\mathrm{max}}$.  In this situation, the soliton oscillated in size.  In Fig.~\ref{fig:oscillateenergy}, we again see that energy is conserved to one part in $10^4$ for a step factor of 1.  For other simulations where the mass was less than $M_{\mathrm{max}}$, the solitons also oscillated.  Similarly, for other simulations where the mass was greater than $M_{\mathrm{max}}$, the soliton collapsed.  These results were consistent with results that appear in~\cite{Chavanis2016}.

Here, $M$ is the soliton's mass in terms of the maximum soliton mass defined in Eq.~(\ref{eq:Mmax}).  The best fit dynamical time for both graphs was 0.786. This is longer than the calculated dynamical time of 0.675 code units.  The difference in these values is unlikely to be caused by a lack of spatial resolution because there was a negligible difference between the oscillation frequencies when the resolution was 256 versus 512.   Therefore we believe that the cause is deformation of the Gaussian profile over time, into a more realistic profile.  We know this occurs because we fit a Gaussian to the density profile and looked at how the R-squared value changes over time. The value started at exactly 1 and evolved to approximately 0.998.  This was also the value that the exact solution took. Since we know that the change in the profile has an effect on the maximum mass a soliton can have before collapsing~\cite{Chavanis2016,Eby2018}, it is reasonable to think that the oscillation frequencies might be scaled differently for the different profiles.  We verified that this is the likely explanation by analytically calculating the dynamical time for the exact solution and found that the dynamical time would be 0.791 code units. This new value matches well with the dynamical time found through the simulations.  This also suggests the results of this section are robust beyond the Gaussian ansatz.

\subsection{Exploding solitons}

According to~\cite{Chavanis2016}, under certain initial conditions, we expect to see exploding solitons which are solitons that grow without bound.  Specifically, when the soliton mass is below the critical mass $M_c = \frac{\sqrt{3}}{2}M_{\mathrm{max}}$ and the radius of the soliton in less than $R = \frac{1+\sqrt{1-(\frac{M}{M_c})^2}}{2M}$ but greater than the radius that maximizes the effective potential, explosions should occur. This value of the radius is the number of equilibrium radii of a soliton with mass $M_{\mathrm{max}}$ [i.e. $R \equiv R_{eq}(M_{\mathrm{max}})$].  Because the Gaussian ansatz is only an approximate solution, the criteria for observing different phenomena, such as collapsing solitons and oscillating solitons, changes slightly.  For instance, the maximum mass for a soliton with a Gaussian form is $M_{\mathrm{max}} = 1.085\frac{\hbar}{\sqrt{Gm\abs{a_s}}}$ instead of the exact value of $M_{\mathrm{max}} = 1.012\frac{\hbar}{\sqrt{Gm\abs{a_s}}}$~\cite{Chavanis2016}.

\begin{figure*}[!tbp]
  \centering
  \begin{minipage}[t]{0.49\textwidth}
  \textbf{NO SI WITH NO PHASE SHIFT}
    \begin{tabular}{cc}
    \includegraphics[ width=.36\textwidth]{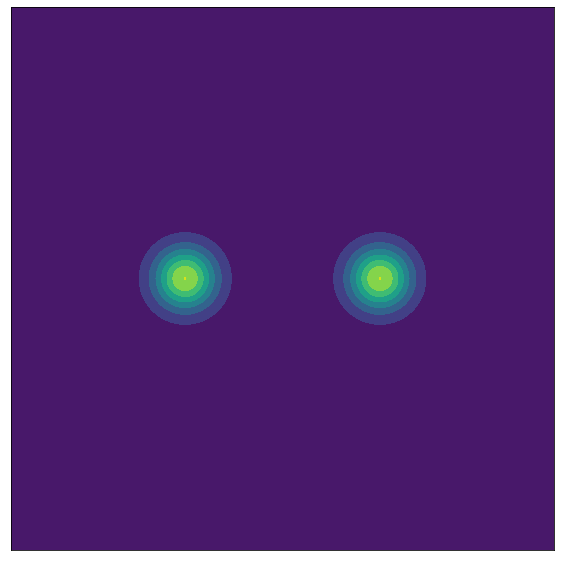}&   \includegraphics[ width=.36\textwidth]{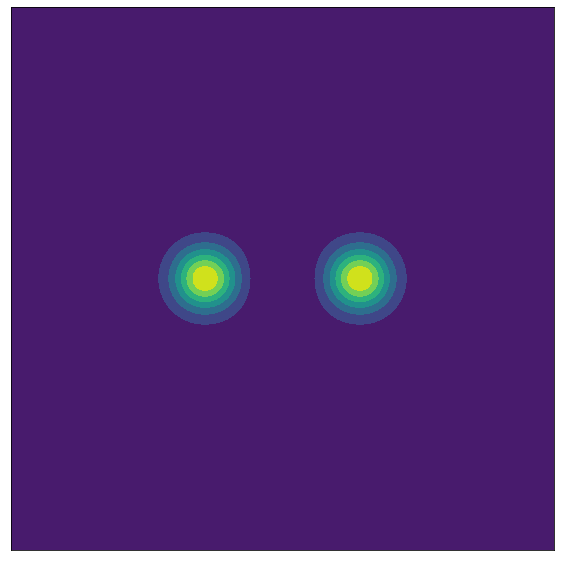} \\
    (a) t = 0.01 & (b) t = 0.02 \\[6pt]
     \includegraphics[ width=.36\textwidth]{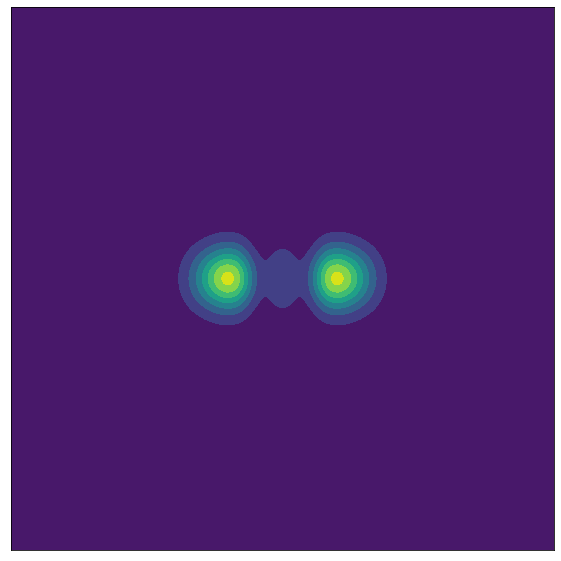} &   \includegraphics[ width=.36\textwidth]{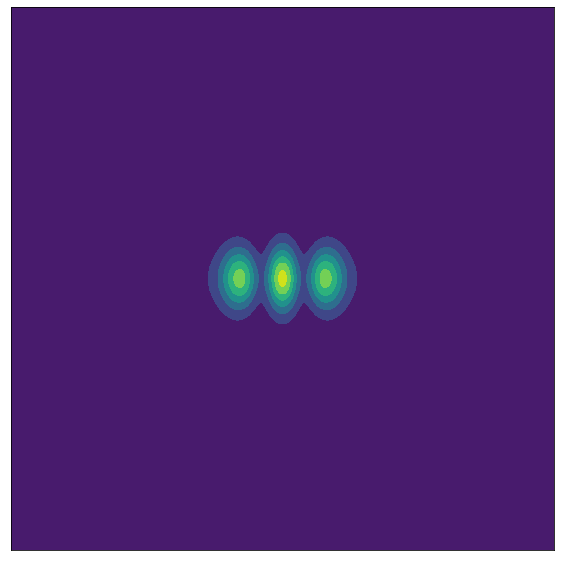} \\
    (c) t = 0.03 & (d) t = 0.04 \\[6pt]
    \includegraphics[ width=.36\textwidth]{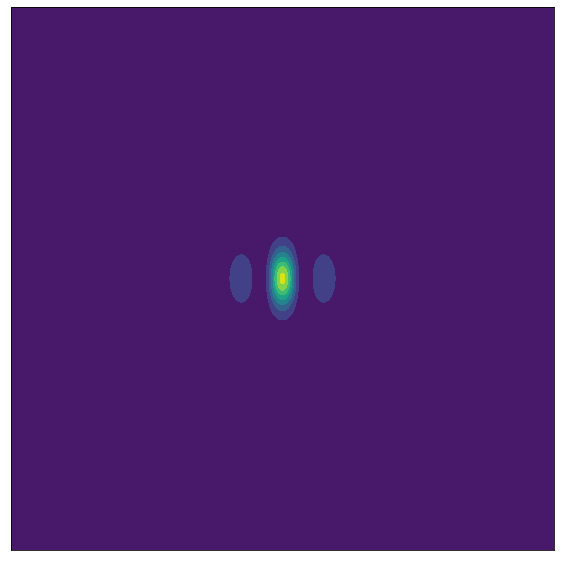} &   \includegraphics[ width=.36\textwidth]{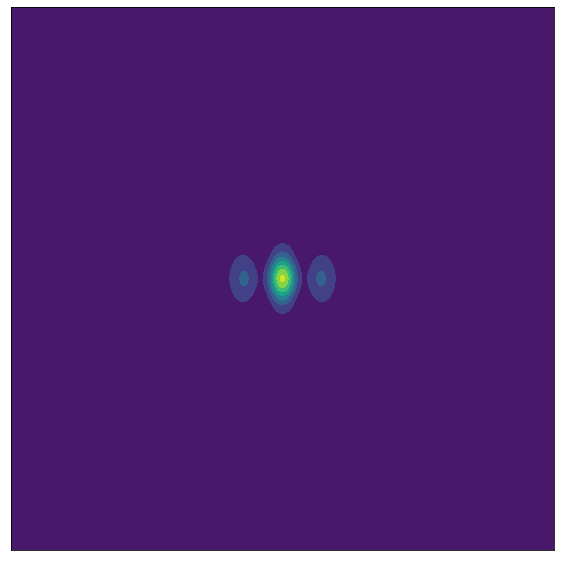} \\
    (e) t = 0.05 & (f) t = 0.06 \\[6pt]
    \includegraphics[ width=.36\textwidth]{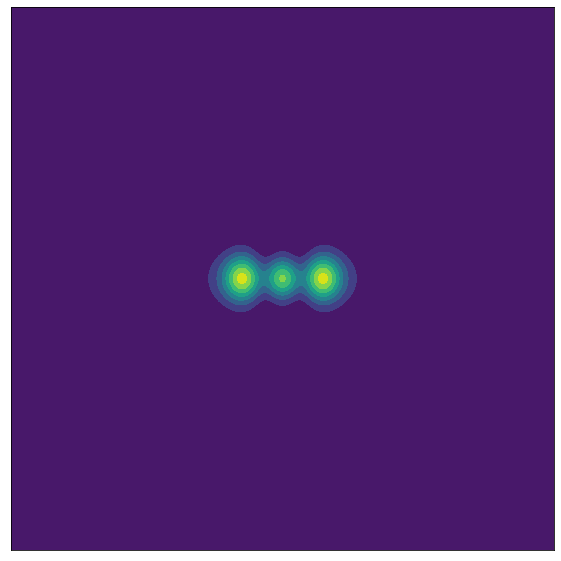} &   \includegraphics[ width=.36\textwidth]{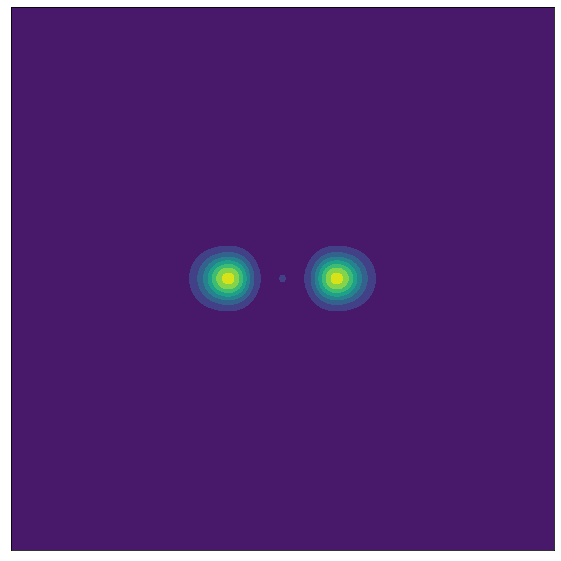} \\
    (g) t = 0.07 & (h) t = 0.08 \\[6pt]
    \includegraphics[ width=.36\textwidth]{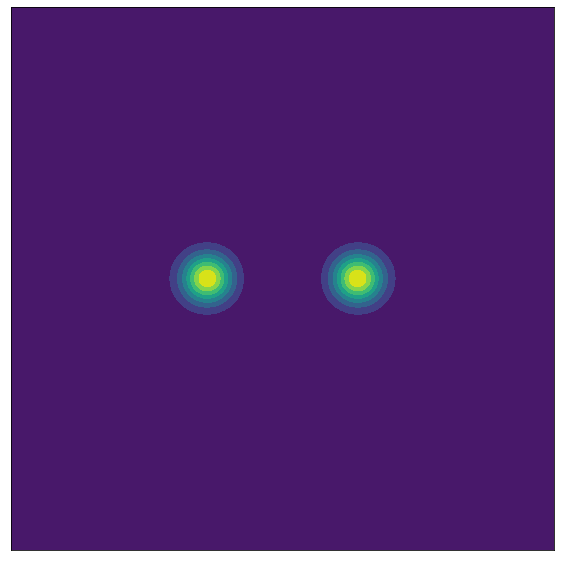} &   \includegraphics[ width=.36\textwidth]{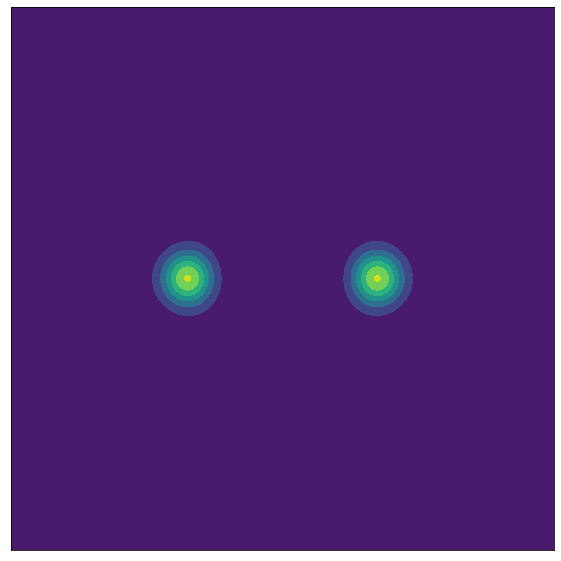} \\
    (i) t = 0.09 & (j) t = 0.10 \\[6pt]
    \end{tabular}
    \caption{Two colliding solitons with no phase shift and no self-interaction. These plots show contours of constant density. Time progresses across each row left to right, and the time under each frame is indicated in code units. The total duration of the simulation is 7.6~Gyr. The two solitons have the same shape and size from when they leave the collision area and when they enter.  Each soliton passes though the other unaffected.}
    \label{fig:fig2_1}	
  \end{minipage}
  \hfill
  \begin{minipage}[t]{0.49\textwidth}
  \textbf{ATTRACTIVE SI WITH NO PHASE SHIFT}
    \begin{tabular}{cc}
    \includegraphics[ width=.36\textwidth]{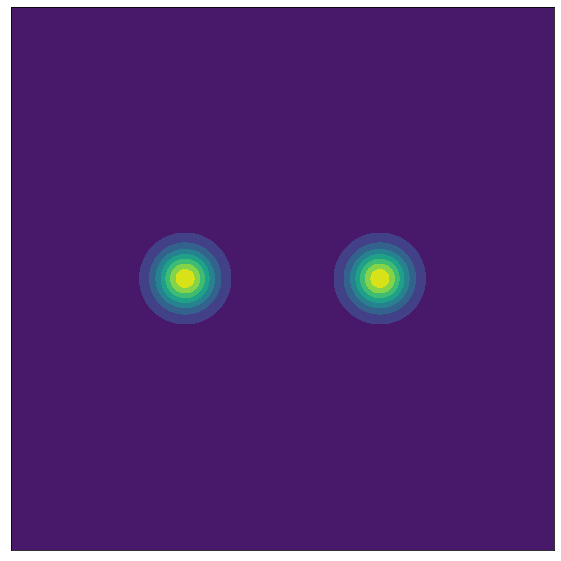}&   \includegraphics[ width=.36\textwidth]{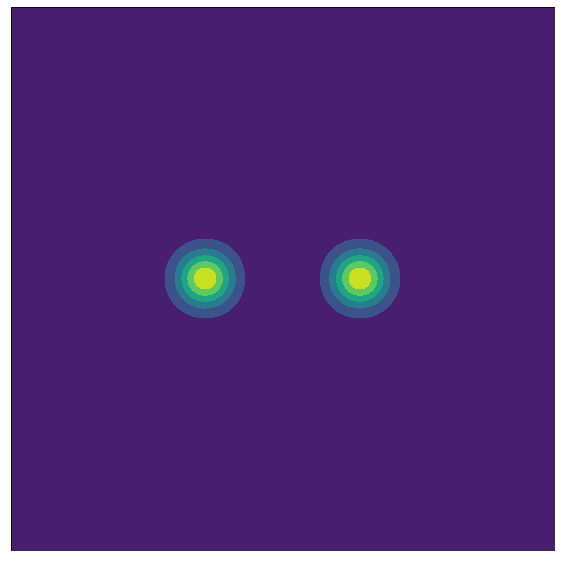} \\
    (a) t = 0.01 & (b) t = 0.02 \\[6pt]
    \includegraphics[ width=.36\textwidth]{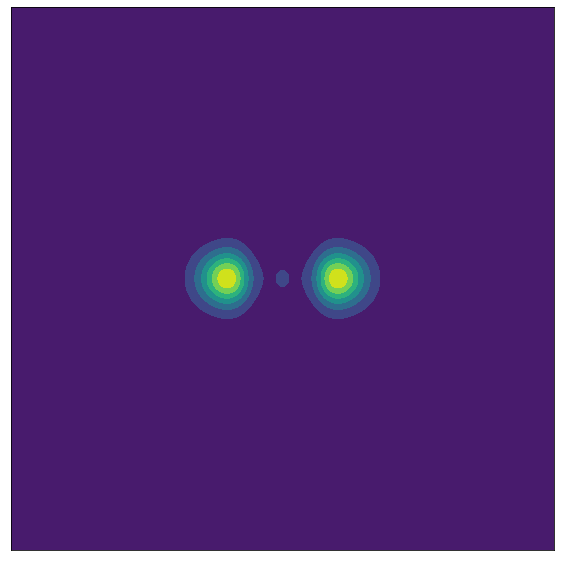} &   \includegraphics[ width=.36\textwidth]{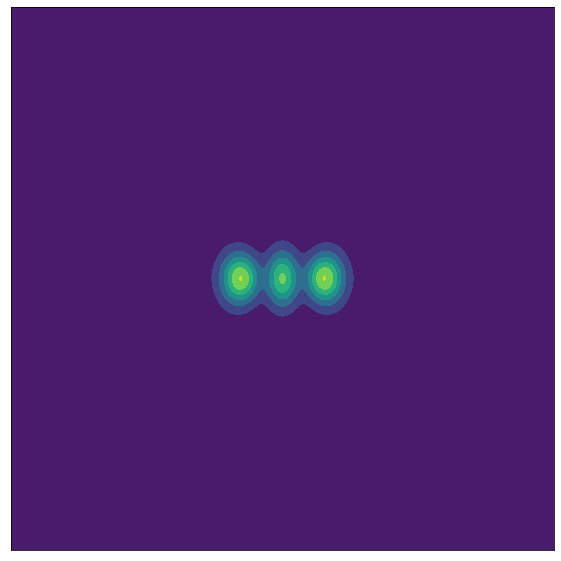} \\
    (c) t = 0.03 & (d) t = 0.04 \\[6pt]
    \includegraphics[ width=.36\textwidth]{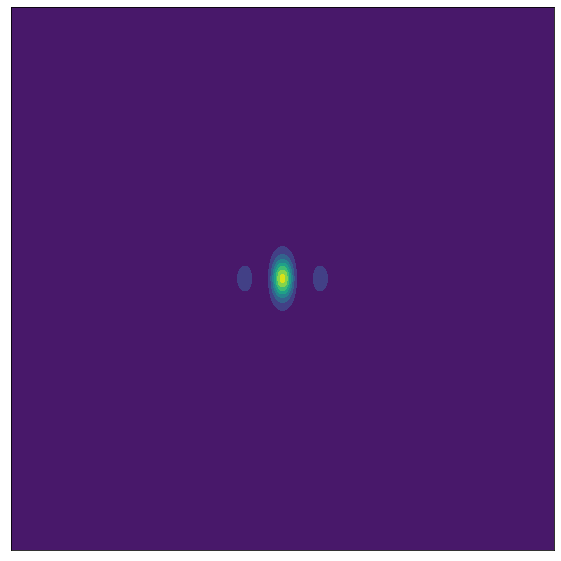} &   \includegraphics[ width=.36\textwidth]{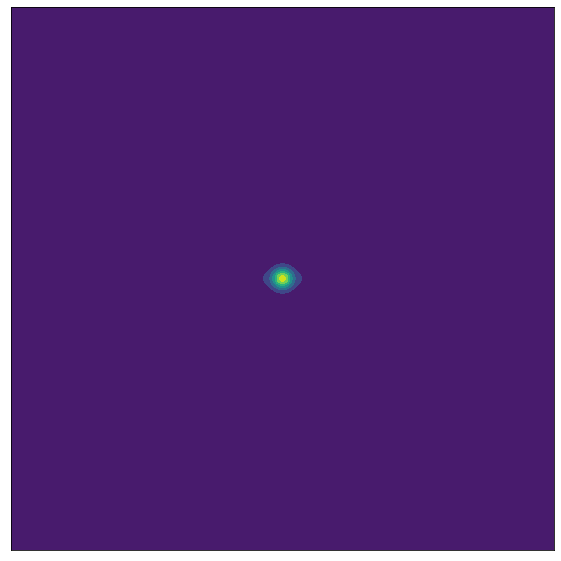} \\
    (e) t = 0.05 & (f) t = 0.06 \\[6pt]
    \includegraphics[ width=.36\textwidth]{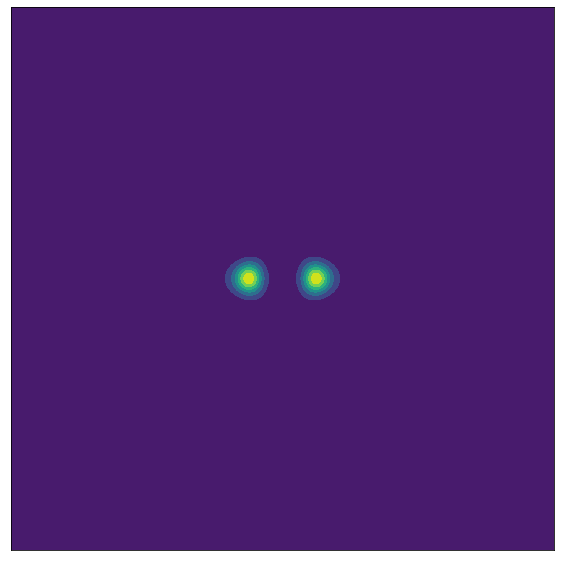} &   \includegraphics[ width=.36\textwidth]{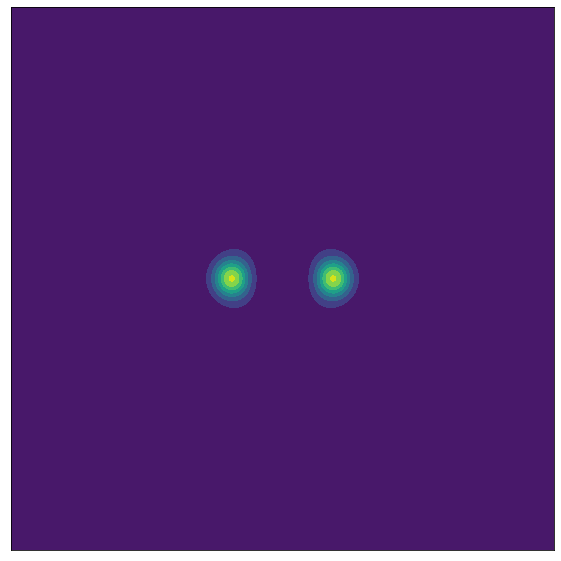} \\
    (g) t = 0.07 & (h) t = 0.08 \\[6pt]
    \includegraphics[ width=.36\textwidth]{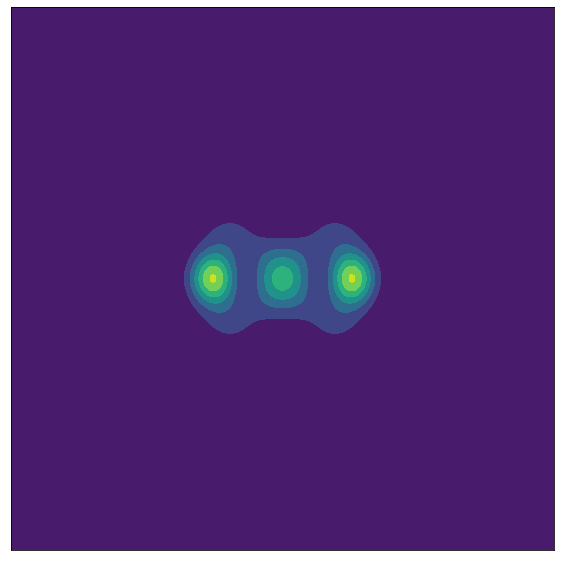} &   \includegraphics[ width=.36\textwidth]{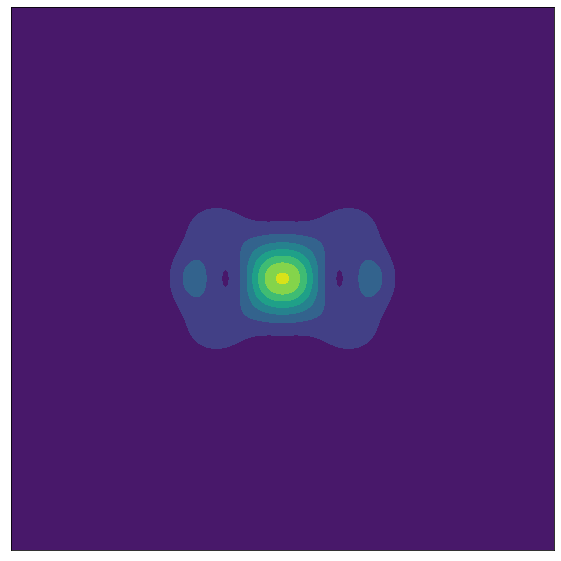} \\
    (i) t = 0.09 & (j) t = 0.10 \\[6pt]
    \end{tabular}
    \caption{Two colliding solitons with no phase shift with an attractive self interaction. These plots show contours of constant density. Time progresses across each row left to right, and the time under each frame is indicated in code units. The duration for the simulation is 7.6~Gyr.  $\kappa = -0.02$ or $\lambda = -9.7 \times 10^{-91}$.  The inclusion of an attractive self interaction causes the solitons to become distorted.  With relatively small masses, the solitons do not collapse into a black hole, and instead merge and oscillate in size.}
    \label{fig:fig2_1_kap}	
  \end{minipage}
\end{figure*}

\begin{figure*}[!tbp]
  \centering
  \begin{minipage}[t]{0.49\textwidth}
  \textbf{REPULSIVE SI WITH NO PHASE SHIFT}
    \begin{tabular}{cc}
    \includegraphics[ width=.36\textwidth]{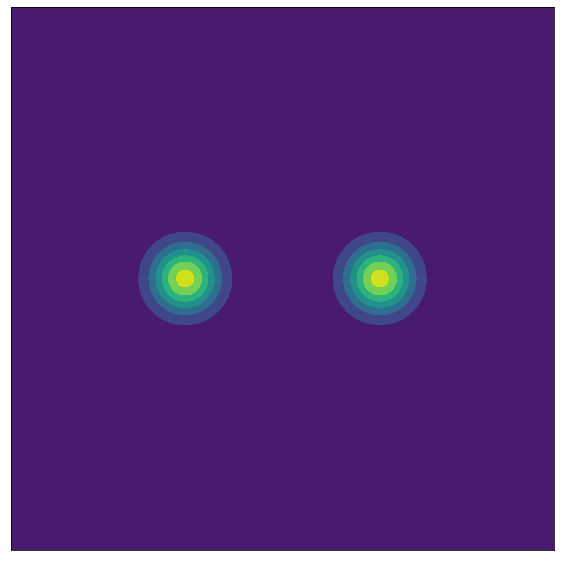}&   \includegraphics[ width=.36\textwidth]{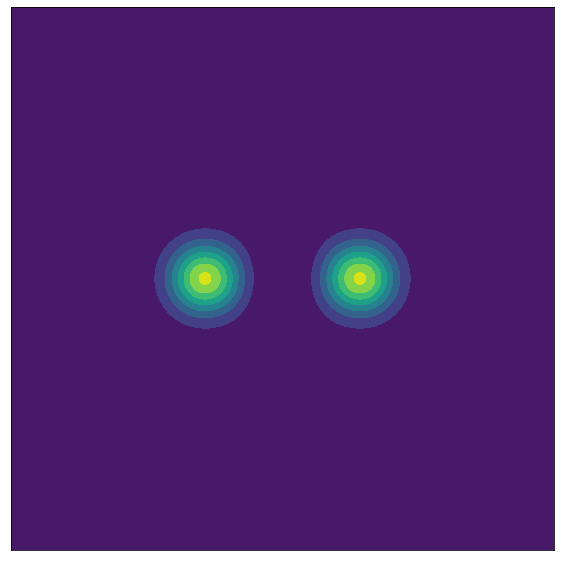} \\ 
    (a) t = 0.01 & (b) t = 0.02 \\[6pt]
    \includegraphics[ width=.36\textwidth]{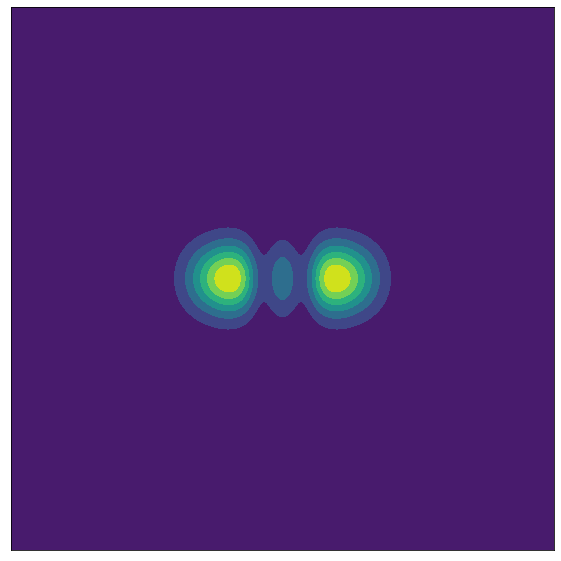} &   \includegraphics[ width=.36\textwidth]{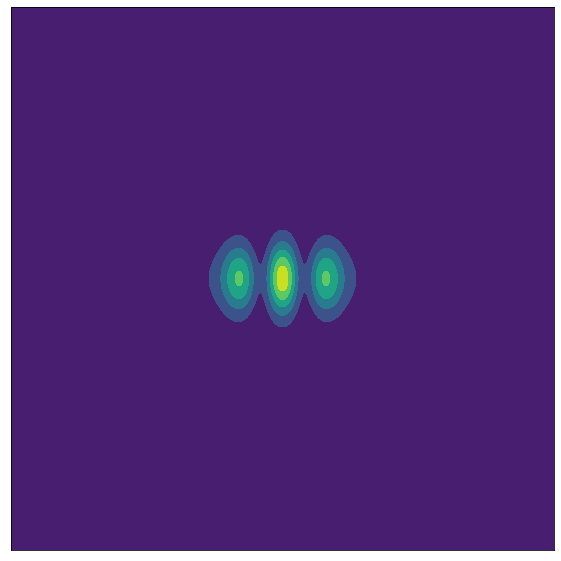} \\
    (c) t = 0.03 & (d) t = 0.04 \\[6pt]
    \includegraphics[ width=.36\textwidth]{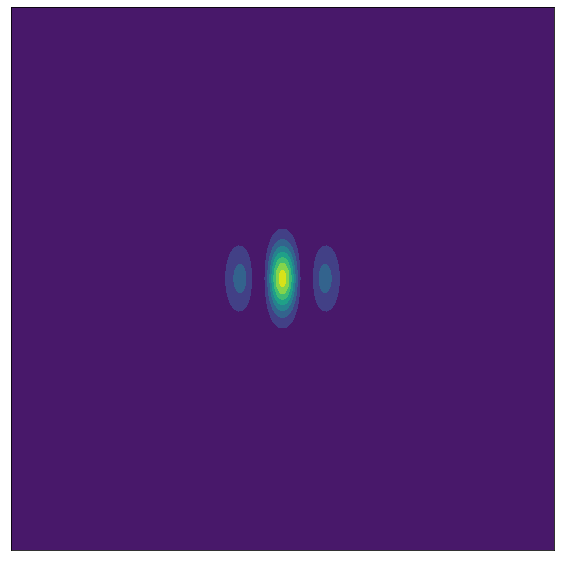} &   \includegraphics[width=.36\textwidth]{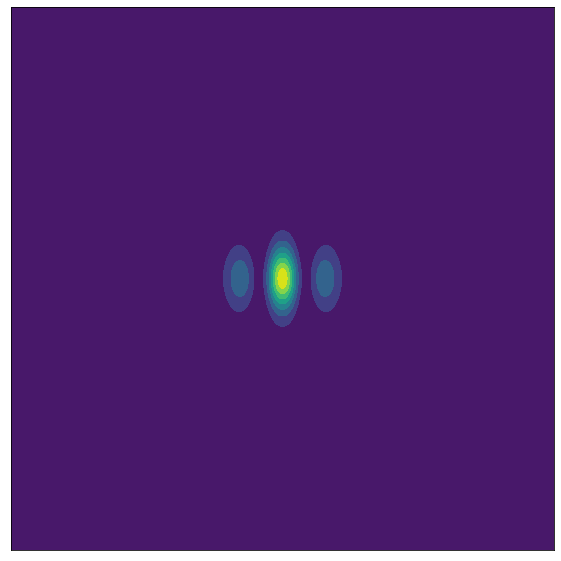} \\
    (e) t = 0.05 & (f) t = 0.06 \\[6pt]
    \includegraphics[ width=.36\textwidth]{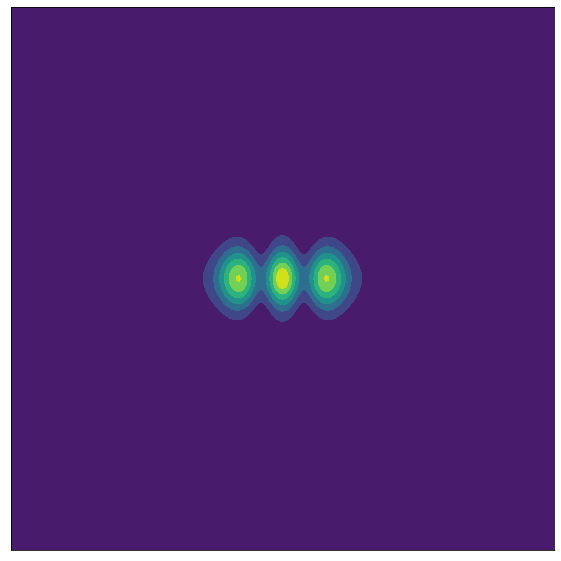} &   \includegraphics[ width=.36\textwidth]{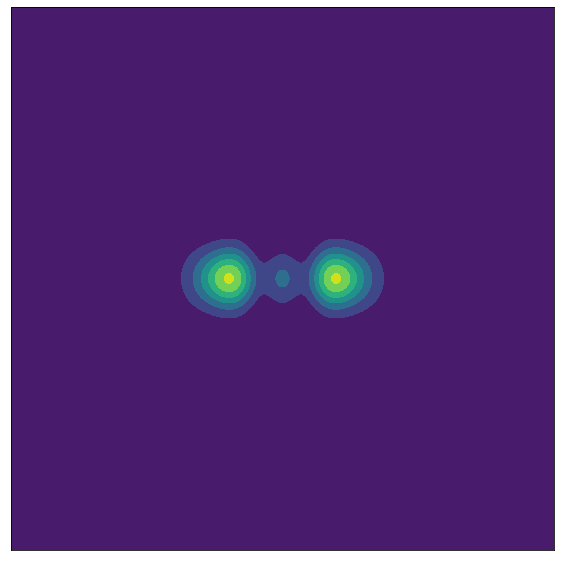} \\
    (g) t = 0.07 & (h) t = 0.08 \\[6pt]
    \includegraphics[ width=.36\textwidth]{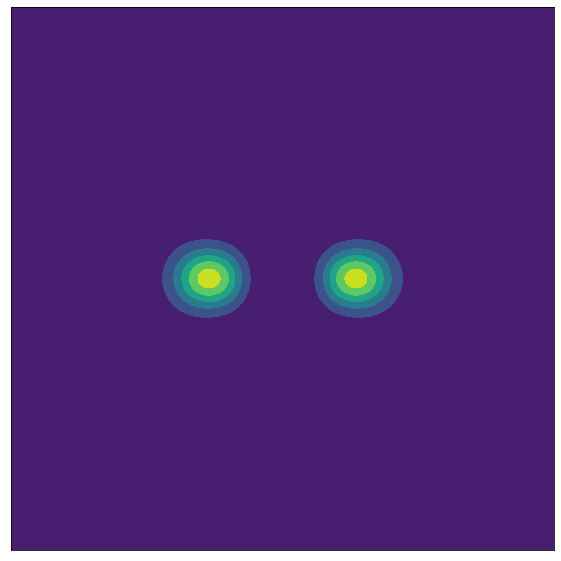} &   
    \includegraphics[ width=.36\textwidth]{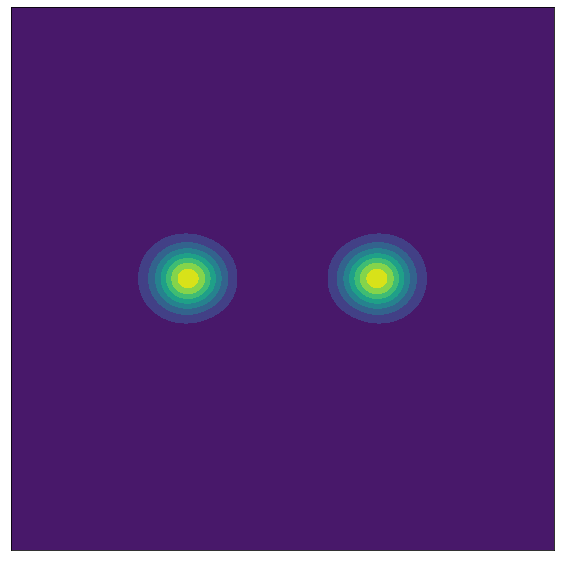} \\
    (i) t = 0.09 & (j) t = 0.10 \\[6pt]
    \end{tabular}
    \caption{Two colliding solitons with no phase shift with a repulsive self-interaction. These plots show contours of constant density. Time progresses across each row left to right, and the time under each frame is indicated in code units.  The duration is 7.6~Gyr.  $\kappa = 0.02$ or $\lambda = 9.7 \times 10^{-91}$.  The results are similar to that in Fig.~\ref{fig:fig2_1} but with the resultant solitons slightly enlarged.  There are also minor differences in the intermediate time steps.}
    \label{fig:fig2_1_kap_rep}
  \end{minipage}
  \hfill
  \begin{minipage}[t]{0.49\textwidth}
  \textbf{NO SI WITH PHASE SHIFT}
    \begin{tabular}{cc}
    \includegraphics[ width=.36\textwidth]{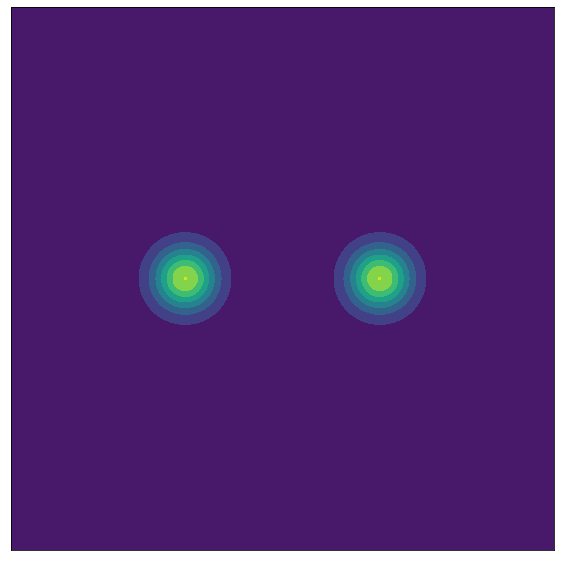}&   \includegraphics[ width=.36\textwidth]{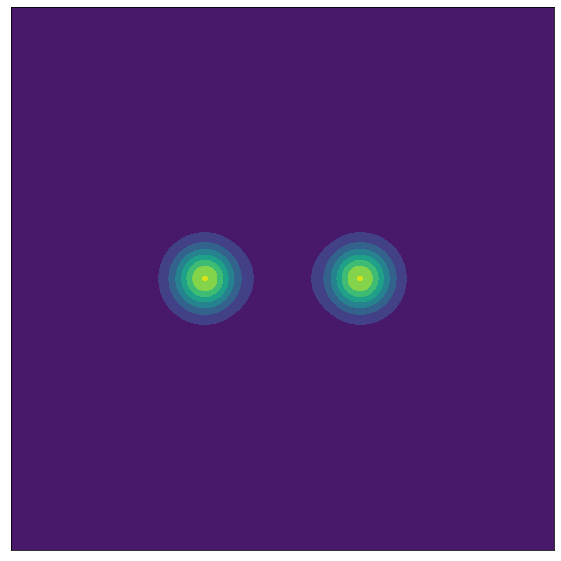} \\
    (a) t = 0.01 & (b) t = 0.02 \\[6pt]
    \includegraphics[ width=.36\textwidth]{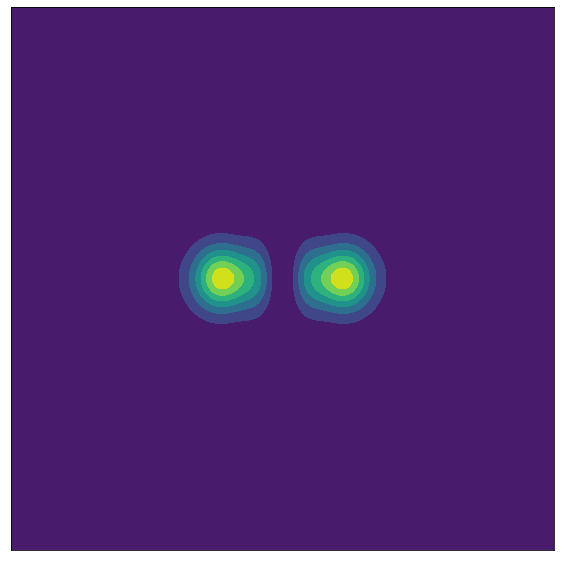} &   \includegraphics[ width=.36\textwidth]{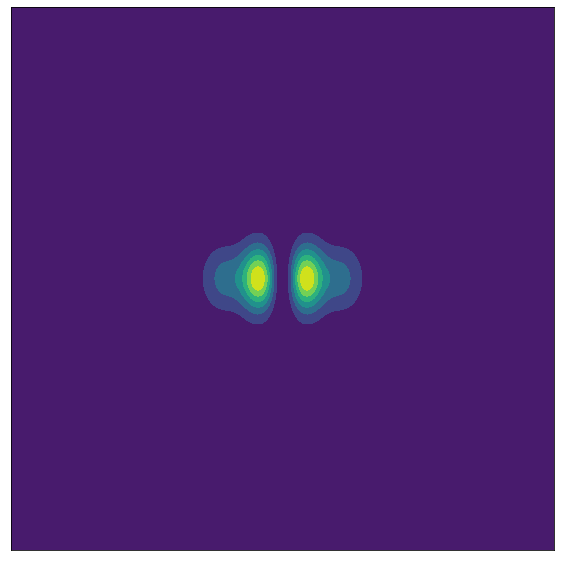} \\
    (c) t = 0.03 & (d) t = 0.04 \\[6pt]
    \includegraphics[ width=.36\textwidth]{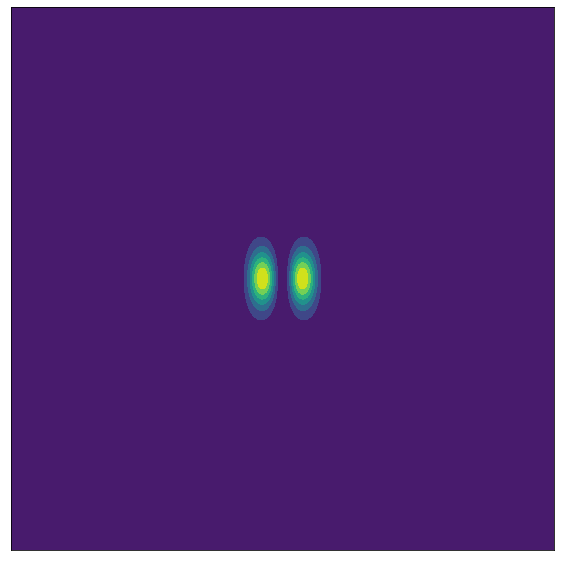} &   \includegraphics[ width=.36\textwidth]{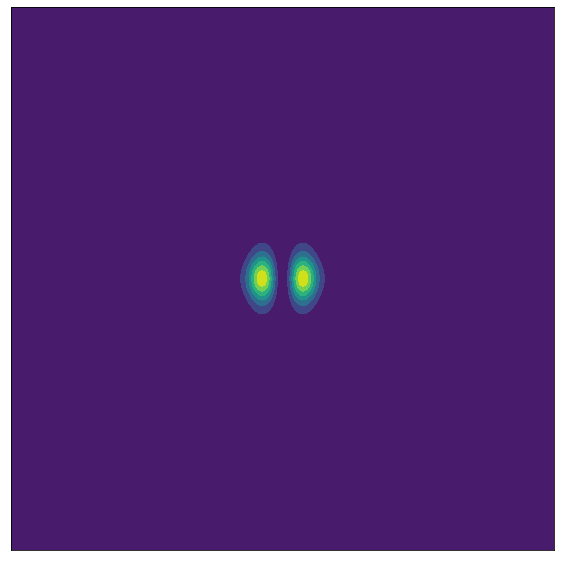} \\
    (e) t = 0.05 & (f) t = 0.06 \\[6pt]
    \includegraphics[ width=.36\textwidth]{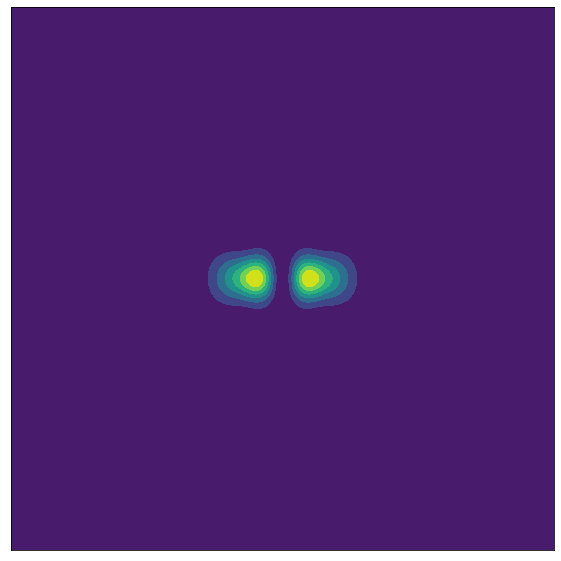} &   \includegraphics[ width=.36\textwidth]{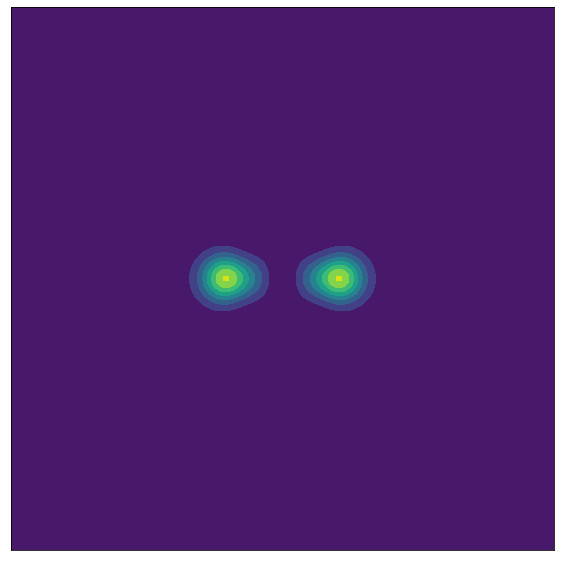} \\
    (g) t = 0.07 & (h) t = 0.08 \\[6pt]
    \includegraphics[ width=.36\textwidth]{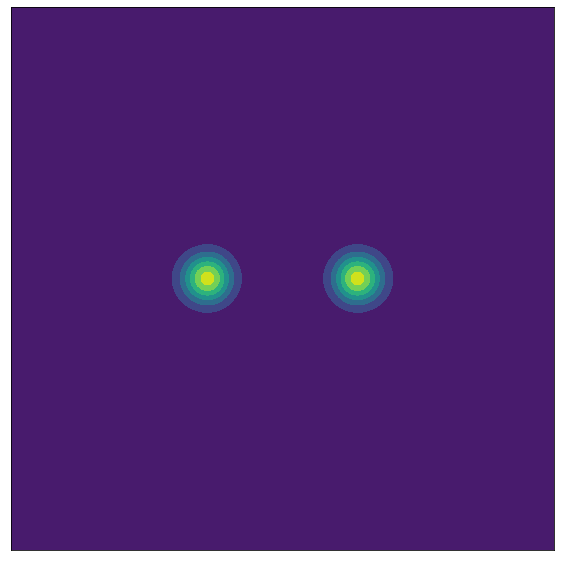} &   \includegraphics[ width=.36\textwidth]{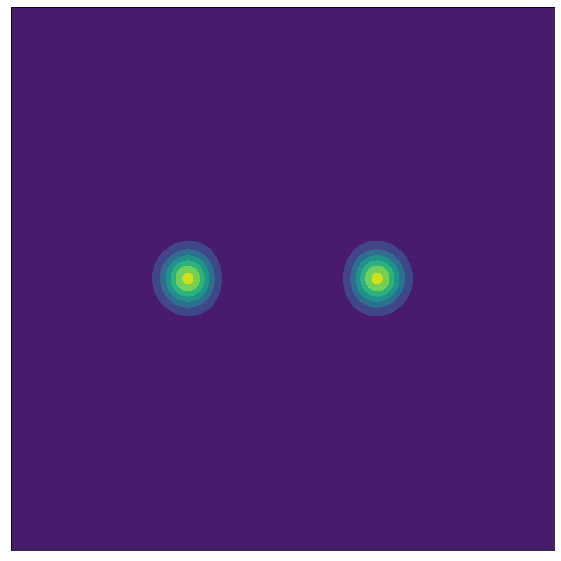} \\
    (i) t = 0.09 & (j) t = 0.10 \\[6pt]
    \end{tabular}
    \caption{Two colliding solitons with a phase shift of pi and no self-interaction. These plots show contours of constant density. Time progresses across each row left to right, and the time under each frame is indicated in code units. The total duration of the simulation is 7.6~Gyr.  The effective repulsive force from the phase difference causes the solitons to decelerate and then move in opposite directions.  The initial and final soliton profiles are the same.}
    \label{fig:fig2_2}	
  \end{minipage}
\end{figure*}

\begin{figure*}[!tbp]
  \centering
  \begin{minipage}[t]{0.49\textwidth}
  \textbf{ATTRACTIVE SI WITH PHASE SHIFT}
    \begin{tabular}{cc}
    \includegraphics[ width=.36\textwidth]{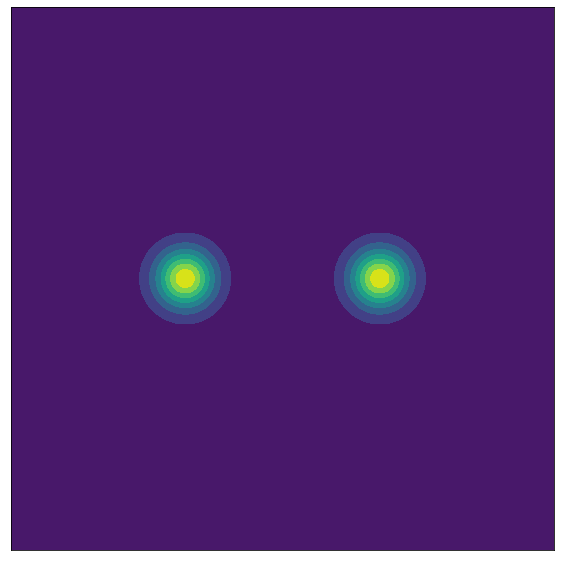}&   \includegraphics[ width=.36\textwidth]{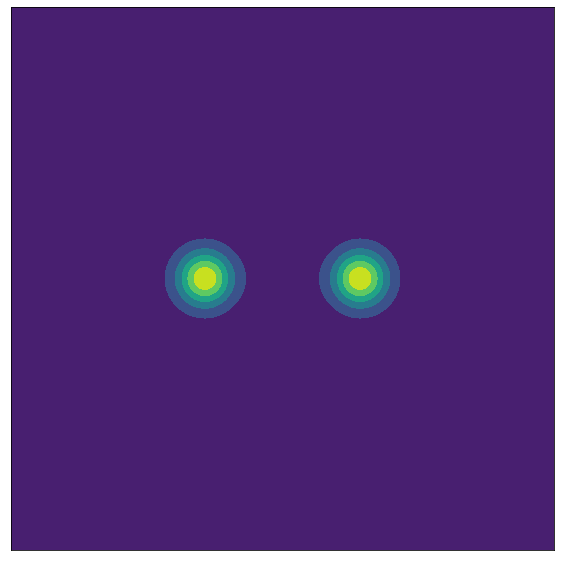} \\
    (a) t = 0.01 & (b) t = 0.02 \\[6pt]
     \includegraphics[ width=.36\textwidth]{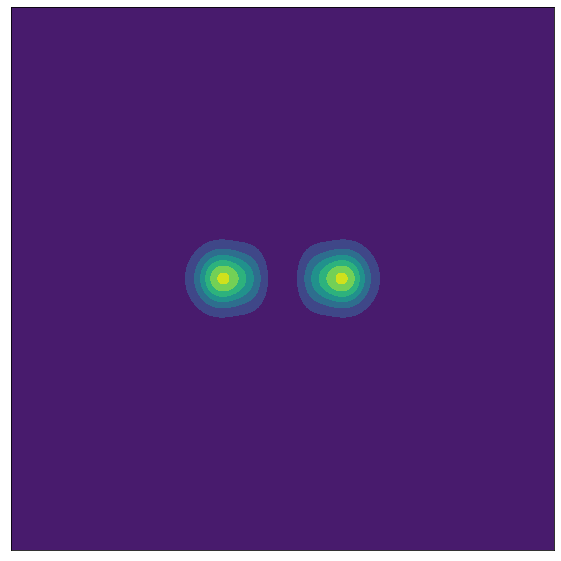} &   \includegraphics[ width=.36\textwidth]{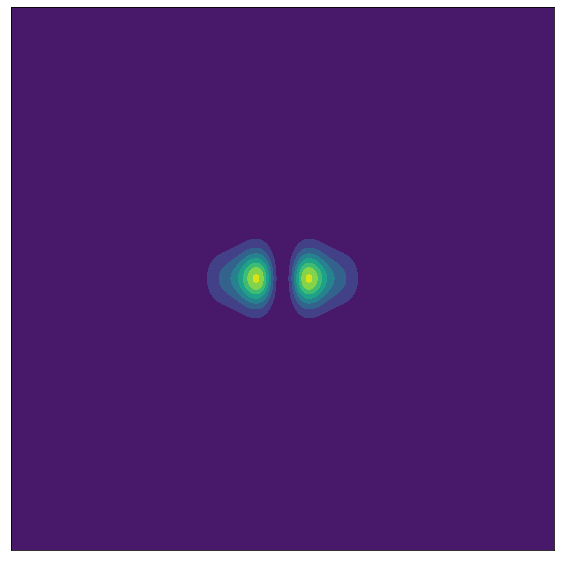} \\
    (c) t = 0.03 & (d) t = 0.04 \\[6pt]
    \includegraphics[ width=.36\textwidth]{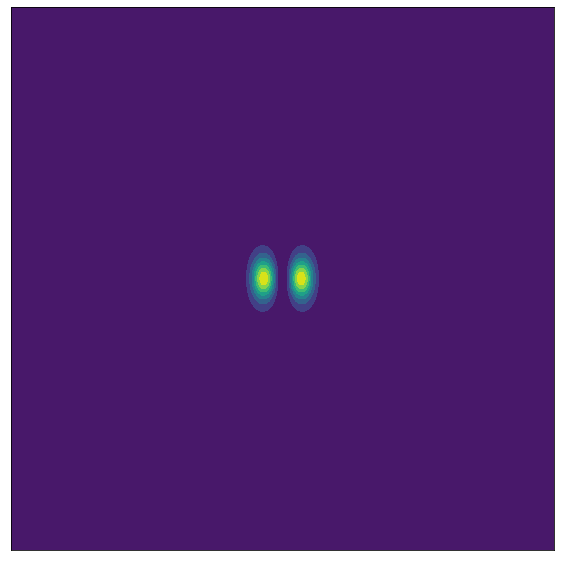} &   \includegraphics[ width=.36\textwidth]{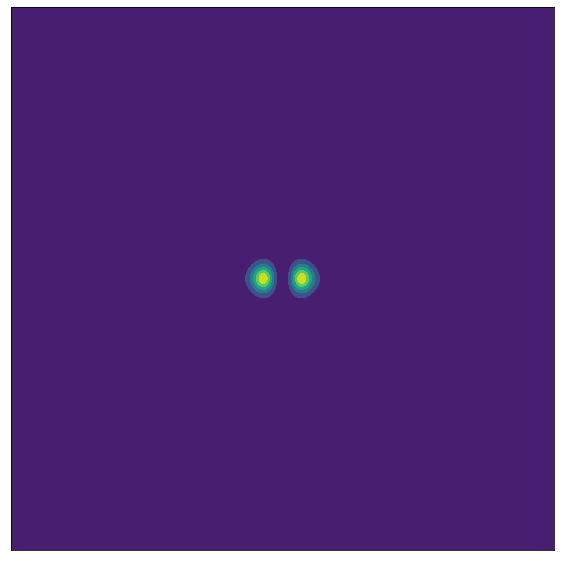} \\
    (e) t = 0.05 & (f) t = 0.06 \\[6pt]
    \includegraphics[ width=.36\textwidth]{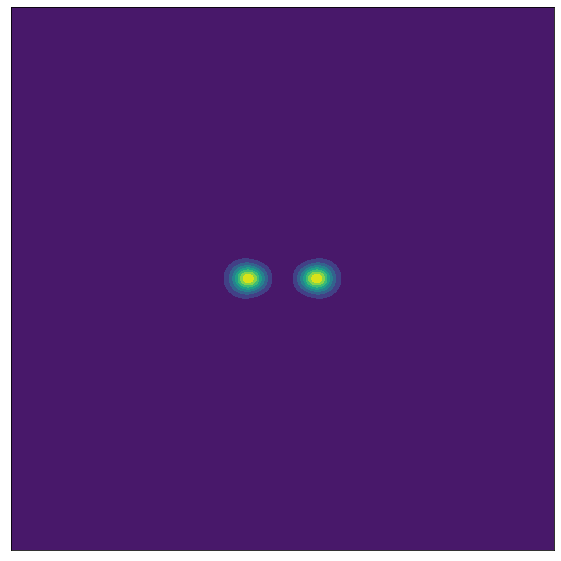} &   \includegraphics[ width=.36\textwidth]{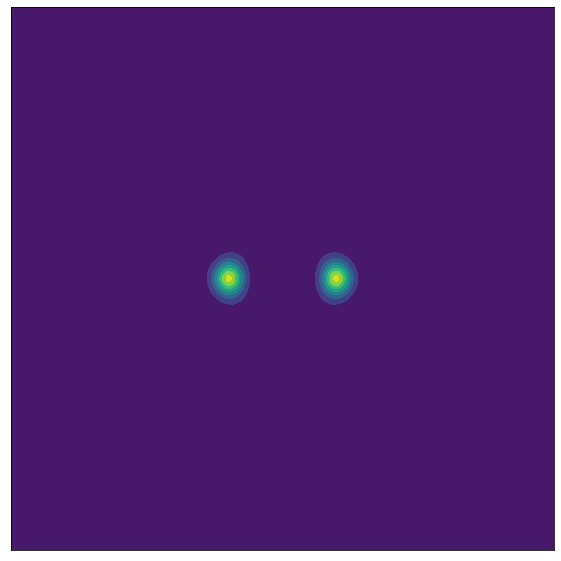} \\
    (g) t = 0.07 & (h) t = 0.08 \\[6pt]
    \includegraphics[ width=.36\textwidth]{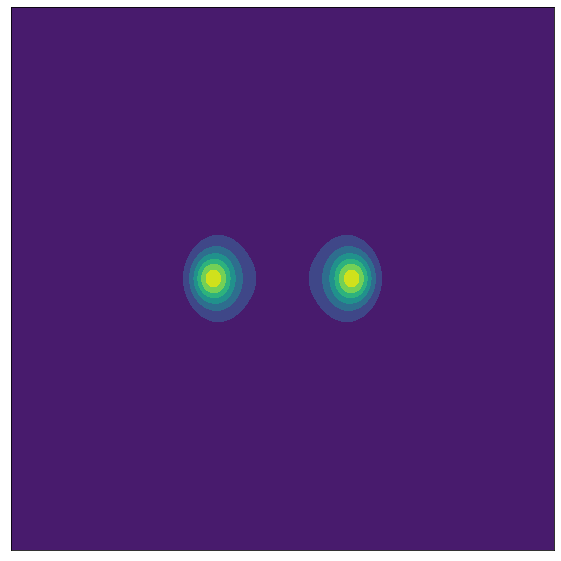} &   \includegraphics[ width=.36\textwidth]{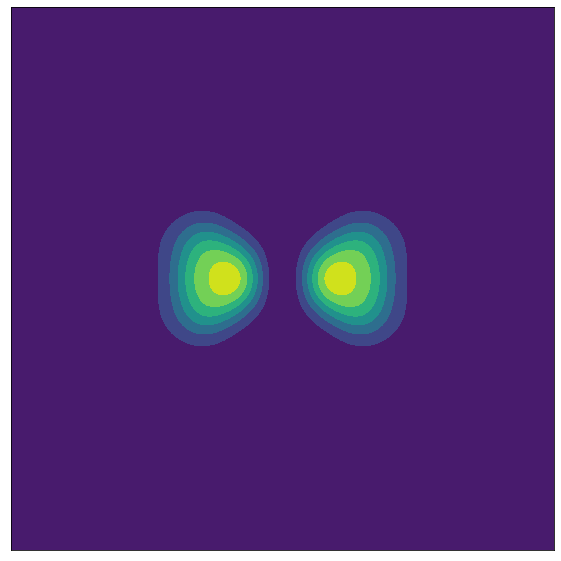} \\
    (i) t = 0.09 & (j) t = 0.10 \\[6pt]
    \end{tabular}
    \caption{Two colliding solitons with a phase shift of pi with an attractive self-interaction. These plots show contours of constant density. Time progresses across each row left to right, and the time under each frame is indicated in code units. The total duration is 7.6~Gyr. $\kappa = -0.02$ or $\lambda = -9.7 \times 10^{-91}$.The repulsive force caused by the phase shift is strong enough to repel the solitons, however, the attractive self-interaction still causes the solitons to oscillate in size.}
    \label{fig:fig2_2_kap}
  \end{minipage}
  \hfill
  \begin{minipage}[t]{0.49\textwidth}
  \textbf{REPULSIVE SI WITH PHASE SHIFT}
    \begin{tabular}{cc}
    \includegraphics[ width=.36\textwidth]{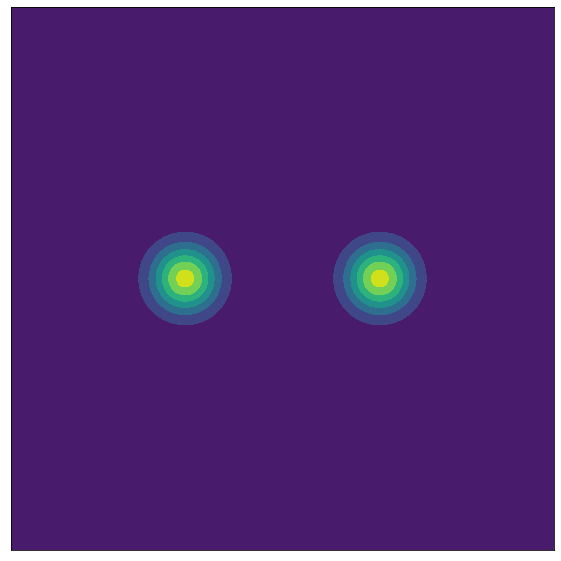}&   \includegraphics[width=.36\textwidth]{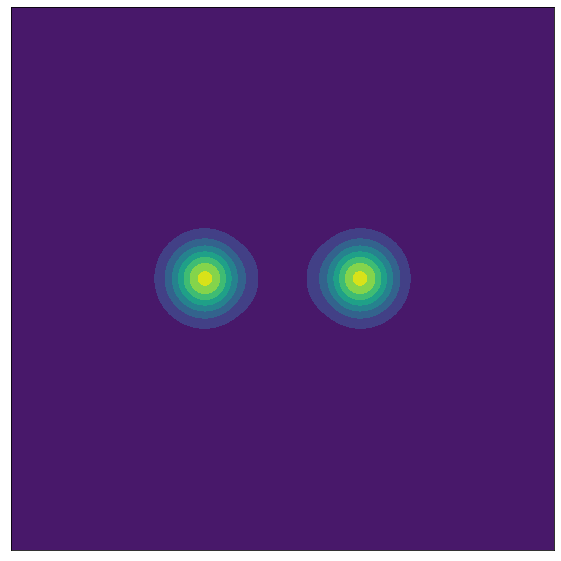} \\
    (a) t = 0.01 & (b) t = 0.02 \\[6pt]
    \includegraphics[ width=.36\textwidth]{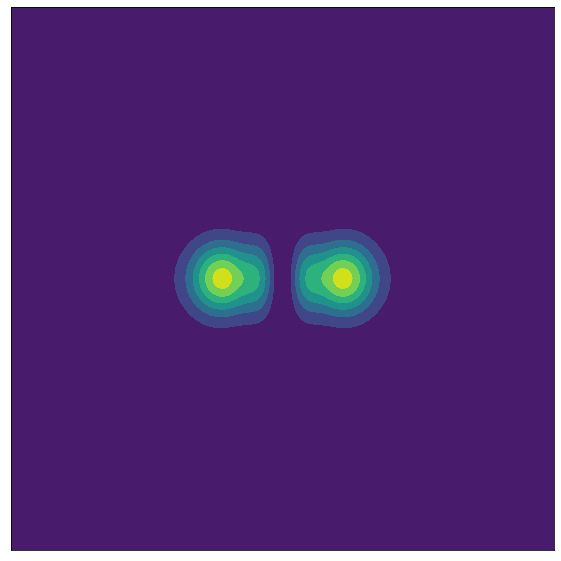} &   \includegraphics[width=.36\textwidth]{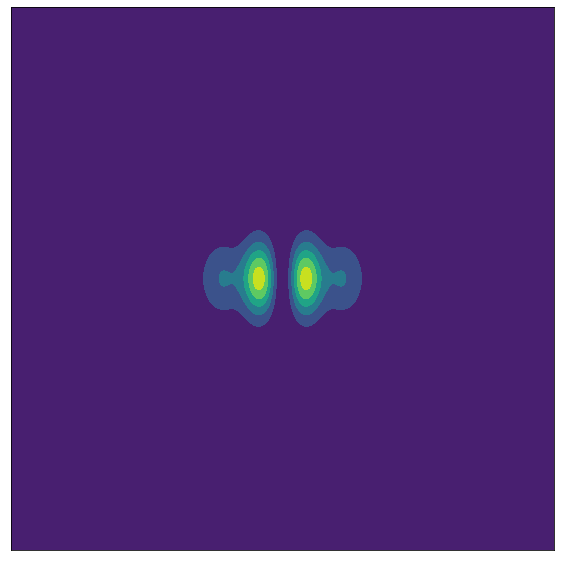} \\
    (c) t = 0.03 & (d) t = 0.04 \\[6pt]
    \includegraphics[ width=.36\textwidth]{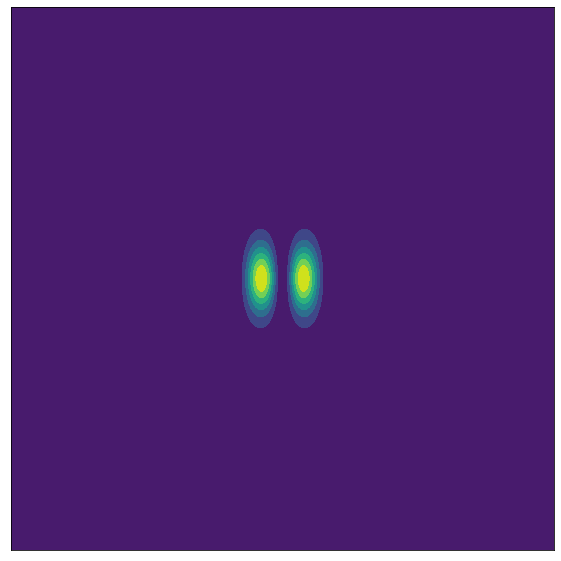} &   \includegraphics[width=.36\textwidth]{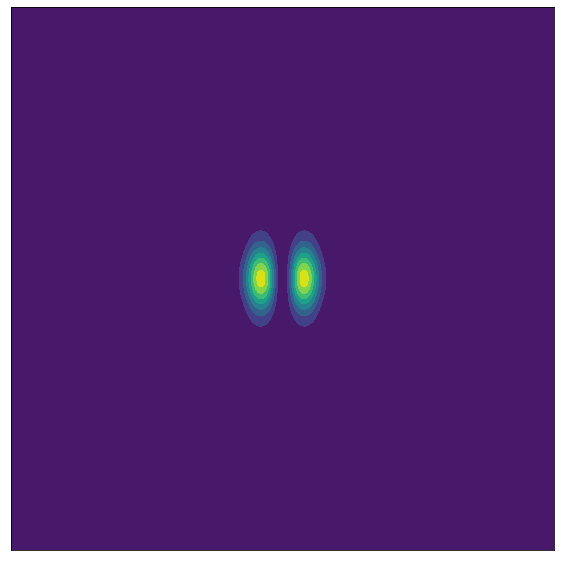} \\
    (e) t = 0.05 & (f) t = 0.06 \\[6pt]
    \includegraphics[ width=.36\textwidth]{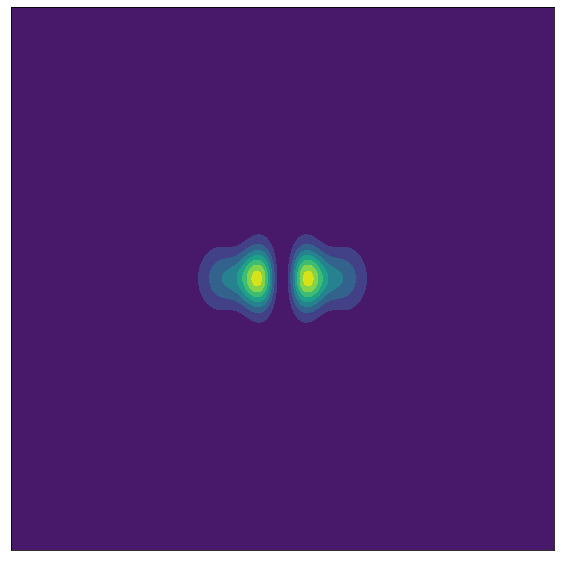} &   \includegraphics[width=.36\textwidth]{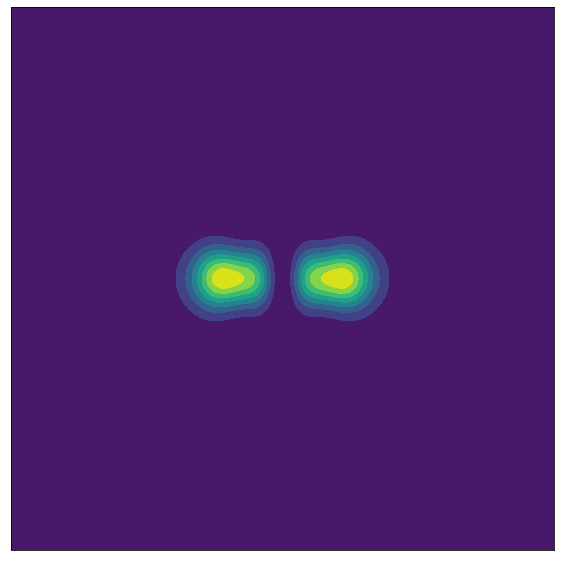} \\
    (g) t = 0.07 & (h) t = 0.08 \\[6pt]
    \includegraphics[ width=.36\textwidth]{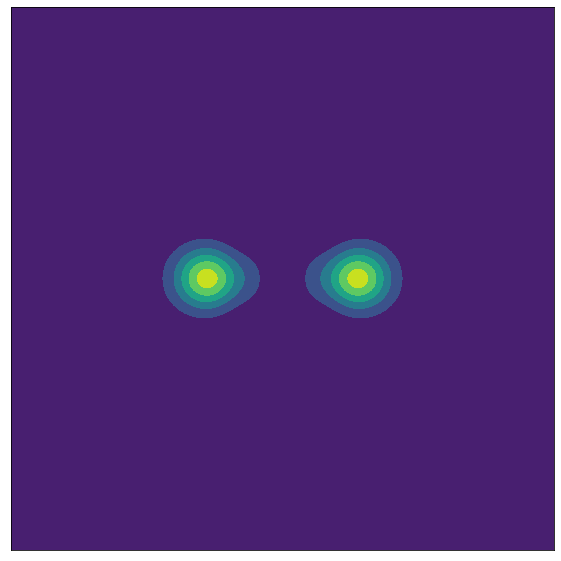} &   
    \includegraphics[ width=.36\textwidth]{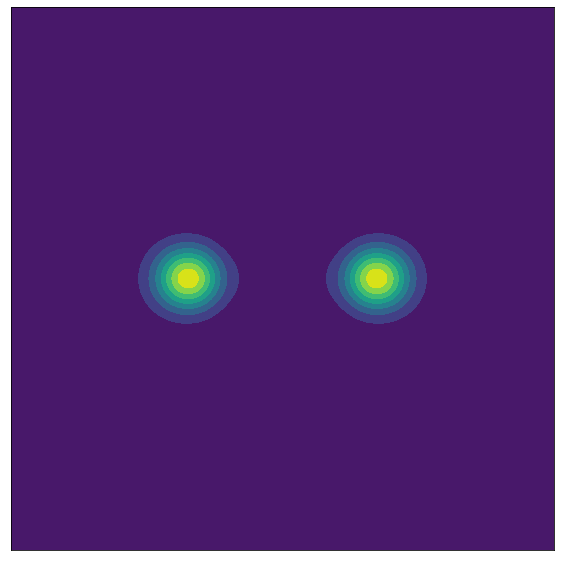} \\
    (i) t = 0.09 & (j) t = 0.10 \\[6pt]
    \end{tabular}
    \caption{Two colliding solitons with a phase shift of pi with a repulsive self-interaction. These plots show contours of constant density. Time progresses across each row left to right, and the time under each frame is indicated in code units. The duration is 7.6~Gyr.  $\kappa = 0.02$ or $\lambda = 9.7 \times 10^{-91}$.  The results are similar to that in Fig.~\ref{fig:fig2_2}.}
    \label{fig:fig2_2_kap_rep}
  \end{minipage}
\end{figure*} 

In Fig.~\ref{fig:explode}, we simulate a soliton with attractive self-interactions.  The initial size is smaller than the critical radius. The mass is also below the critical mass.  The soliton expands in size without collapsing again or oscillating in size.  We find that with a sufficiently small radius, provided that the mass is less than a critical mass, the soliton collapses.

\subsection{Collapsing solitons}
Moving beyond the Gaussian ansatz, we also set up a scenario involving a single soliton with an attractive self-interaction and varied the mass of the soliton. This provides a test of how successful the code produces results without assuming a Gaussian ansatz.  For these simulations, we find agreement with the numerical prediction of the exact maximum mass criteria found in~\cite{Chavanis2011}.  Using $\kappa = -2.0$ and an axion mass of $m = 1 \times 10^{-22}~\mathrm{eV}/\mathrm{c}^2$ (which corresponds to $a_s = -7.6 \times 10^{-60}~\mathrm{fm}$), the maximum mass of a soliton from Eq.~(\ref{eq:maxmass}) is $M_{\mathrm{max}} = 5.66\times 10^{6} M_\odot$ or 2.46 code units.  The starting simulation had the soliton mass at 2.5 code units which is larger than $M_{\mathrm{max}}$.  We found that the soliton collapsed into a singularity which we define as when at least $0.25\%$ of the total mass is contained in one unit cell.  \textsc{PySiUltraLight} can only handle a grid-point to grid-point phase difference of up to $\pi /2$ which limits how dense the collapsed soliton can get. Figure~\ref{fig:collapseenergy} also shows the components of the system's energy over time.  From this, we see that the total system energy is conserved up until the end of the collapse.  For most of the simulation, energy is conserved to better than one part in $10^3$.  After the soliton has collapsed, the total energy change is about 7 percent.  This simulation was done with a step factor of 0.05 which is a measure of the temporal resolution.  We found that decreasing the step factor greatly reduced the percent energy change when there are collapsing solitons.


\section{Multiple Soliton Behavior and Central Potential Examples in PySiUltraLight}
\label{msolitonic}

\textsc{PySiUltralight}, like its predecessor, is capable of simulating multiple solitons simultaneously.  This is useful for studying how solitons merge and interact with one another. The setup that we consider in this section, of multiple interacting solitons goes beyond those explored in~\cite{Chavanis2016}. Unlike the work where we were making direct comparisons with that paper, here we do not need to assume as prior an approximate shape of the soliton density profile and do not use the Gaussian ansatz. The first scenario we set up with multiple solitons is a binary soliton collision.  The examples we use are idealized in that they are head-on collisions and the solitons are of equal mass, however, understanding binary collisions is important for understanding mergers.  In Figs.~\ref{fig:fig2_1} through~\ref{fig:fig2_2_kap}, two solitons move towards each other.  The setup differed only by a phase shift between solitons and the value of the self-coupling.  

In all the figures, the initial setup has two solitons, each of 20 code mass units ($4.6\times 10^7~\mathrm{M}_\odot$), spaced 1.2 code units ($45.6~\mathrm{kpc})$ apart moving towards each other with a relative velocity of 20 code units ($9.8~\mathrm{km/s}$).  The duration in all the simulations is 0.1 code units which is about $7.6~\mathrm{Gyr}$.  We chose this duration so that we would be able to compare our simulation results with those in~\cite{Edwards2018}.  In the scenarios where there is an attractive self-interaction, $\kappa = -0.02$ corresponding to $\lambda = -9.7 \times 10^{-91}$.  When there are repulsive self-interactions, $\kappa = 0.02$.  This corresponds to $\lambda = 9.7 \times 10^{-91}$.  In these, the decay constant is $f_a \approx 10^{14}~\mathrm{GeV}$.  In these figures, the color scales are different in each frame in the same way that the scales are different in Fig.~\ref{fig:explode}.  These figures are useful schematically in understanding the behavior of these collisions.  

\begin{figure}
\begin{tabular}{cc}	
\includegraphics[trim=240 40 240 40, clip, width=.33\textwidth]{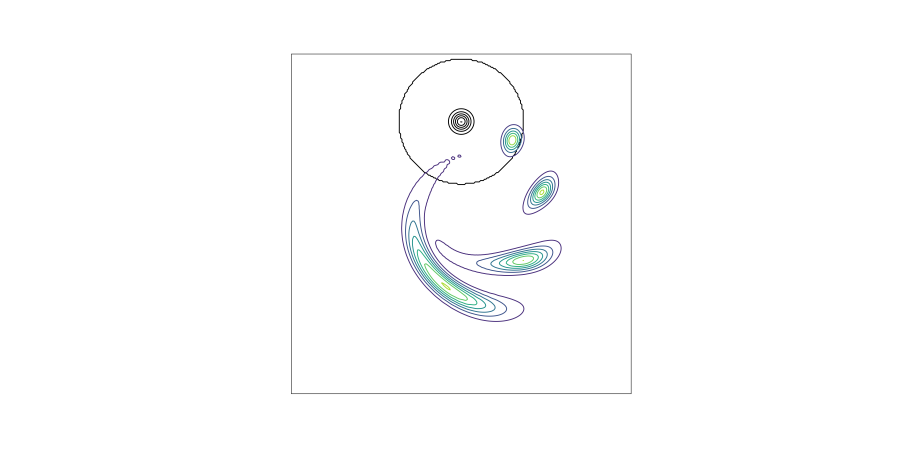}\\
(a) No self-interactions\\
\includegraphics[trim=240 40 240 40, clip, width=.33\textwidth]{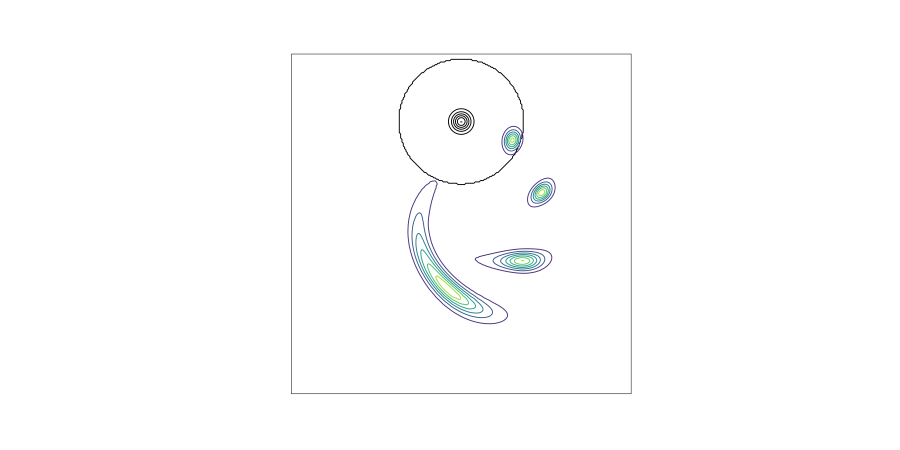}\\
(b) Attractive self-interactions\\
\includegraphics[trim=240 40 240 40, clip, width=.33\textwidth]{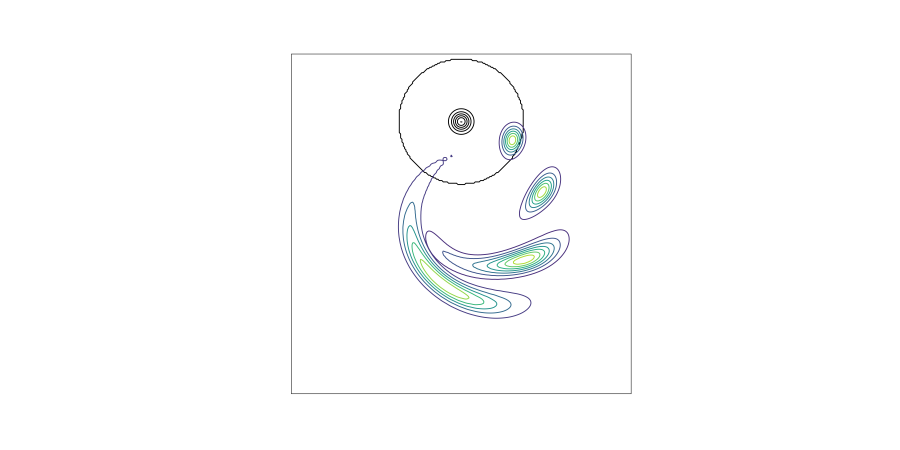}\\
(c) Repulsive self-interactions
\end{tabular}
\caption{Plot (a) shows how a soliton behaves when rotating around a central potential at different times when there are no self-interactions.  Plot (b) has the same initial conditions except it has a coupling of $\kappa = -0.05$.  The most noticeable difference between the two plots is that the soliton with an attractive self-interaction is less spread out by the central potential.  Plot (c) has a repulsive coupling of $\kappa = 0.05$.  Here, we see that the soliton is more spread out than the simulations with no coupling or with an attractive coupling.}
\label{fig:orbit}
\end{figure}

In Figs.~\ref{fig:fig2_1},~\ref{fig:fig2_1_kap}, and~\ref{fig:fig2_1_kap_rep}, there is no phase shift between solitons.  The difference between these figures is that in Fig.~\ref{fig:fig2_1} the solitons enter and leave the collision area unaltered while in Fig.~\ref{fig:fig2_1_kap} they are distorted.  The solitons pass through each other unaffected which is to be expected when there is no self-interaction term.  However, when there is an attractive self-interaction term, the solitons merge and then oscillate in size.  There are only minor differences between Figs.~\ref{fig:fig2_1} and~\ref{fig:fig2_1_kap_rep}.  The biggest difference is that with the repulsive self-interaction, the resultant solitons are slightly larger in size.   In Figs.~\ref{fig:fig2_2},~\ref{fig:fig2_2_kap}, and~\ref{fig:fig2_2_kap_rep} there is a phase shift of $\pi$ between the two solitons.  This phase shift creates an effective repulsive force between the solitons~\cite{Paredes2015}.  As with the first three simulations, the main difference is how the solitons enter and exit the collision zone.

In~\cite{Edwards2018}, the authors run a simulation in \textsc{PyUltraLight} where a soliton rotates around a central potential.  We recreated this and then included an attractive self-interaction to see what differences exist with this inclusion of a self-coupling.  For the setup, an axion mass of $m = 1 \times 10^{-22}~\mathrm{eV}/\mathrm{c}^2$ was used, the box length was ten code units ($380~\mathrm{kpc})$, the duration was 0.4 code units ($30.2~\mathrm{Gyr}$), the central mass was 1000 code units ($2.3\times 10^9~\mathrm{M}_\odot$), the soliton mass was 12 code units ($2.8\times 10^7~\mathrm{M}_\odot$), the starting distance away from the central mass was three code units ($114~\mathrm{kpc}$), and the initial tangential velocity was 16 code units ($7.9~\mathrm{km/s}$).  Figure~\ref{fig:orbit} shows three scenarios of a soliton orbiting around a central potential, one without self-interactions, one with an attractive self-interaction, and one with repulsive self-interactions.  In this figure the colorful lines represent constant density contours.  As the soliton rotates around the central mass, the soliton stretches as the parts of the soliton closer to the center rotate faster than the outside portions.  When we include an attractive self-interaction, we can see the soliton is less disrupted than the case when there were no self-interactions.  Simulations with repulsive self-interactions had the opposite effect in that the solitons became more spread out.  These observations are important because they suggest self-interactions play a role in determining the lifetimes of solitons.  It is important to note contours with the same colors at different times do not necessarily correspond to the same density.  This is because each figure is made from overlaying different snapshots in time.


\section{Conclusions}
\label{conclusions}

In this paper, we provide an outline for altering \textsc{PyUltraLight} to include self-interactions.  We provided background information on the physics of ultralight dark matter and detailed the changes we made to \textsc{PyUltraLight}.  We tested the altered program's integrity by comparing the new version with the original version and by confirming energy was still conserved in the simulations.  Using this altered version, we attempted to verify predictions made in~\cite{Chavanis2016} which quantified phenomena that one expects to see when ultralight dark matter has attractive self-interactions.  The oscillation frequency results from our code behaved similarly to analytic predictions but had a different dynamical time.  This discrepancy is likely caused by the approximate solution of the Gaussian profile being adjusted as the simulation progressed.  We looked at the conditions needed to see exploding solitons by using an approximate Gaussian soliton profile.  When the soliton mass was less than the critical mass and the starting soliton size was sufficiently small, we saw the soliton explode. The predictions for finding a maximum mass for a soliton with attractive self-interactions matched well. 

We also ran simulations that compared situations with attractive self-interactions, repulsive self-interactions, and without self-interactions, including binary soliton collisions and solitons rotating around a central potential.  Qualitatively, there were noticeable differences when self-interactions were introduced in these scenarios.  The attractive self-interactions cause the solitons to more easily stick to one another which is to be expected.  Attractive self-interactions also made it more difficult for the solitons to be tidally disrupted when the solitons rotated around the central potential.  We also found that having repulsive self-interactions had the effect of spreading out the soliton profile more when orbiting a central potential.  This effect may have an impact on the survival times of infalling solitons.

For future work, we would like to implement absorbing boundary conditions using a sponge layer.  Another possible phenomenon to test using this code is the formation of vortices in a BEC which should occur with sufficiently large repulsive self-interactions~\cite{Rindler2011}.

\section*{ACKNOWLEDGMENTS}
\setlength{\parskip}{0pt}
We would like to thank Lisa Bouchard, Annika Peter, Arka Banerjee, Ethan Nadler, and Katelin Schutz for offering comments on this paper.  We would like to thank Adam Dukehart for his insight on how the energy changes with the inclusion of self-interactions.  We would also like to thank Ethan Nadler and J. Luna Zagorac for their helpful discussions about our simulations results. We thank Eli Levinson-Falk, Richard Easther, Brian Nord, Nathan Musoke, Anthony Mirasola, Kay Kirkpatrick, Ed Copeland, Risa Wechsler, and Juna Kollmeier for additional helpful conversations. C. P. W. would like to thank all workers who made this research possible, especially those at the University of New Hampshire (including Michelle Waltz and Katie Makem-Boucher), the Aspen Center for Physics, which is supported by National Science Foundation (NSF) Grant No. PHY-1607611, and the Kavli Institute for Theoretical Physics, where this research was supported in part by the NSF under Grant No. PHY-1748958. Contributions to this project by C. P. W. were also supported by DOE Grant No. DE-SC0020220. This paper honors the memory of Breonna Taylor.

\bibliography{masterlibrary}

\begin{thebibliography}{66}%
\makeatletter
\providecommand \@ifxundefined [1]{%
 \@ifx{#1\undefined}
}%
\providecommand \@ifnum [1]{%
 \ifnum #1\expandafter \@firstoftwo
 \else \expandafter \@secondoftwo
 \fi
}%
\providecommand \@ifx [1]{%
 \ifx #1\expandafter \@firstoftwo
 \else \expandafter \@secondoftwo
 \fi
}%
\providecommand \natexlab [1]{#1}%
\providecommand \enquote  [1]{``#1''}%
\providecommand \bibnamefont  [1]{#1}%
\providecommand \bibfnamefont [1]{#1}%
\providecommand \citenamefont [1]{#1}%
\providecommand \href@noop [0]{\@secondoftwo}%
\providecommand \href [0]{\begingroup \@sanitize@url \@href}%
\providecommand \@href[1]{\@@startlink{#1}\@@href}%
\providecommand \@@href[1]{\endgroup#1\@@endlink}%
\providecommand \@sanitize@url [0]{\catcode `\\12\catcode `\$12\catcode
  `\&12\catcode `\#12\catcode `\^12\catcode `\_12\catcode `\%12\relax}%
\providecommand \@@startlink[1]{}%
\providecommand \@@endlink[0]{}%
\providecommand \url  [0]{\begingroup\@sanitize@url \@url }%
\providecommand \@url [1]{\endgroup\@href {#1}{\urlprefix }}%
\providecommand \urlprefix  [0]{URL }%
\providecommand \Eprint [0]{\href }%
\providecommand \doibase [0]{http://dx.doi.org/}%
\providecommand \selectlanguage [0]{\@gobble}%
\providecommand \bibinfo  [0]{\@secondoftwo}%
\providecommand \bibfield  [0]{\@secondoftwo}%
\providecommand \translation [1]{[#1]}%
\providecommand \BibitemOpen [0]{}%
\providecommand \bibitemStop [0]{}%
\providecommand \bibitemNoStop [0]{.\EOS\space}%
\providecommand \EOS [0]{\spacefactor3000\relax}%
\providecommand \BibitemShut  [1]{\csname bibitem#1\endcsname}%
\let\auto@bib@innerbib\@empty
\bibitem [{\citenamefont {Buckley}\ and\ \citenamefont
  {Peter}(2018)}]{Buckley2018}%
  \BibitemOpen
  \bibfield  {author} {\bibinfo {author} {\bibfnamefont {M.~R.}\ \bibnamefont
  {Buckley}}\ and\ \bibinfo {author} {\bibfnamefont {A.~H.~G.}\ \bibnamefont
  {Peter}},\ }\href {\doibase 10.1016/j.physrep.2018.07.003} {\bibfield
  {journal} {\bibinfo  {journal} {Phys. Rept.}\ }\textbf {\bibinfo {volume}
  {761}},\ \bibinfo {pages} {1} (\bibinfo {year} {2018})},\ \Eprint
  {http://arxiv.org/abs/1712.06615} {arXiv:1712.06615 [astro-ph.CO]}
  \BibitemShut {NoStop}%
\bibitem [{\citenamefont {Ade}\ \emph {et~al.}(2016)\citenamefont {Ade} \emph
  {et~al.}}]{Ade2015}%
  \BibitemOpen
  \bibfield  {author} {\bibinfo {author} {\bibfnamefont {P.~A.~R.}\
  \bibnamefont {Ade}} \emph {et~al.} (\bibinfo {collaboration} {Planck}),\
  }\href {\doibase 10.1051/0004-6361/201525830} {\bibfield  {journal} {\bibinfo
   {journal} {Astron. Astrophys.}\ }\textbf {\bibinfo {volume} {594}},\
  \bibinfo {pages} {A13} (\bibinfo {year} {2016})},\ \Eprint
  {http://arxiv.org/abs/1502.01589} {arXiv:1502.01589 [astro-ph.CO]}
  \BibitemShut {NoStop}%
\bibitem [{\citenamefont {Schwabe}\ \emph {et~al.}(2016)\citenamefont
  {Schwabe}, \citenamefont {Niemeyer},\ and\ \citenamefont
  {Engels}}]{Schwabe2016}%
  \BibitemOpen
  \bibfield  {author} {\bibinfo {author} {\bibfnamefont {B.}~\bibnamefont
  {Schwabe}}, \bibinfo {author} {\bibfnamefont {J.~C.}\ \bibnamefont
  {Niemeyer}}, \ and\ \bibinfo {author} {\bibfnamefont {J.~F.}\ \bibnamefont
  {Engels}},\ }\href {\doibase 10.1103/PhysRevD.94.043513} {\bibfield
  {journal} {\bibinfo  {journal} {Phys. Rev. D}\ }\textbf {\bibinfo {volume}
  {94}},\ \bibinfo {pages} {043513} (\bibinfo {year} {2016})},\ \Eprint
  {http://arxiv.org/abs/1606.05151} {arXiv:1606.05151 [astro-ph.CO]}
  \BibitemShut {NoStop}%
\bibitem [{\citenamefont {Armendariz-Picon}\ and\ \citenamefont
  {Neelakanta}(2014)}]{Armendariz2014}%
  \BibitemOpen
  \bibfield  {author} {\bibinfo {author} {\bibfnamefont {C.}~\bibnamefont
  {Armendariz-Picon}}\ and\ \bibinfo {author} {\bibfnamefont {J.~T.}\
  \bibnamefont {Neelakanta}},\ }\href {\doibase 10.1088/1475-7516/2014/03/049}
  {\bibfield  {journal} {\bibinfo  {journal} {JCAP}\ }\textbf {\bibinfo
  {volume} {03}},\ \bibinfo {pages} {049} (\bibinfo {year} {2014})},\ \Eprint
  {http://arxiv.org/abs/1309.6971} {arXiv:1309.6971 [astro-ph.CO]} \BibitemShut
  {NoStop}%
\bibitem [{\citenamefont {Giesen}\ \emph {et~al.}(2012)\citenamefont {Giesen},
  \citenamefont {Lesgourgues}, \citenamefont {Audren},\ and\ \citenamefont
  {Ali-Haimoud}}]{Giesen2012}%
  \BibitemOpen
  \bibfield  {author} {\bibinfo {author} {\bibfnamefont {G.}~\bibnamefont
  {Giesen}}, \bibinfo {author} {\bibfnamefont {J.}~\bibnamefont {Lesgourgues}},
  \bibinfo {author} {\bibfnamefont {B.}~\bibnamefont {Audren}}, \ and\ \bibinfo
  {author} {\bibfnamefont {Y.}~\bibnamefont {Ali-Haimoud}},\ }\href {\doibase
  10.1088/1475-7516/2012/12/008} {\bibfield  {journal} {\bibinfo  {journal}
  {JCAP}\ }\textbf {\bibinfo {volume} {12}},\ \bibinfo {pages} {008} (\bibinfo
  {year} {2012})},\ \Eprint {http://arxiv.org/abs/1209.0247} {arXiv:1209.0247
  [astro-ph.CO]} \BibitemShut {NoStop}%
\bibitem [{\citenamefont {Aghanim}\ \emph {et~al.}(2020)\citenamefont {Aghanim}
  \emph {et~al.}}]{Aghanim2020}%
  \BibitemOpen
  \bibfield  {author} {\bibinfo {author} {\bibfnamefont {N.}~\bibnamefont
  {Aghanim}} \emph {et~al.} (\bibinfo {collaboration} {Planck}),\ }\href
  {\doibase 10.1051/0004-6361/201833910} {\bibfield  {journal} {\bibinfo
  {journal} {Astron. Astrophys.}\ }\textbf {\bibinfo {volume} {641}},\ \bibinfo
  {pages} {A6} (\bibinfo {year} {2020})},\ \Eprint
  {http://arxiv.org/abs/1807.06209} {arXiv:1807.06209 [astro-ph.CO]}
  \BibitemShut {NoStop}%
\bibitem [{\citenamefont {Del~Popolo}\ and\ \citenamefont
  {Pace}(2016)}]{Popolo2016}%
  \BibitemOpen
  \bibfield  {author} {\bibinfo {author} {\bibfnamefont {A.}~\bibnamefont
  {Del~Popolo}}\ and\ \bibinfo {author} {\bibfnamefont {F.}~\bibnamefont
  {Pace}},\ }\href {\doibase 10.1007/s10509-016-2820-2} {\bibfield  {journal}
  {\bibinfo  {journal} {Astrophys. Space Sci.}\ }\textbf {\bibinfo {volume}
  {361}},\ \bibinfo {pages} {162} (\bibinfo {year} {2016})},\ \bibinfo {note}
  {[Erratum: Astrophys.Space Sci. 361, 225 (2016)]},\ \Eprint
  {http://arxiv.org/abs/1502.01947} {arXiv:1502.01947 [astro-ph.GA]}
  \BibitemShut {NoStop}%
\bibitem [{\citenamefont {Marsh}(2015)}]{Marsh2015}%
  \BibitemOpen
  \bibfield  {author} {\bibinfo {author} {\bibfnamefont {D.~J.~E.}\
  \bibnamefont {Marsh}},\ }\href {\doibase 10.1103/PhysRevD.91.123520}
  {\bibfield  {journal} {\bibinfo  {journal} {Phys. Rev. D}\ }\textbf {\bibinfo
  {volume} {91}},\ \bibinfo {pages} {123520} (\bibinfo {year} {2015})},\
  \Eprint {http://arxiv.org/abs/1504.00308} {arXiv:1504.00308 [astro-ph.CO]}
  \BibitemShut {NoStop}%
\bibitem [{\citenamefont {Spergel}\ and\ \citenamefont
  {Steinhardt}(2000)}]{Spergel2000}%
  \BibitemOpen
  \bibfield  {author} {\bibinfo {author} {\bibfnamefont {D.~N.}\ \bibnamefont
  {Spergel}}\ and\ \bibinfo {author} {\bibfnamefont {P.~J.}\ \bibnamefont
  {Steinhardt}},\ }\href {\doibase 10.1103/PhysRevLett.84.3760} {\bibfield
  {journal} {\bibinfo  {journal} {Phys. Rev. Lett.}\ }\textbf {\bibinfo
  {volume} {84}},\ \bibinfo {pages} {3760} (\bibinfo {year} {2000})},\ \Eprint
  {http://arxiv.org/abs/astro-ph/9909386} {arXiv:astro-ph/9909386} \BibitemShut
  {NoStop}%
\bibitem [{\citenamefont {Avila-Reese}\ \emph {et~al.}(2001)\citenamefont
  {Avila-Reese}, \citenamefont {Colin}, \citenamefont {Valenzuela},
  \citenamefont {D'Onghia},\ and\ \citenamefont {Firmani}}]{Avila_Reese2001}%
  \BibitemOpen
  \bibfield  {author} {\bibinfo {author} {\bibfnamefont {V.}~\bibnamefont
  {Avila-Reese}}, \bibinfo {author} {\bibfnamefont {P.}~\bibnamefont {Colin}},
  \bibinfo {author} {\bibfnamefont {O.}~\bibnamefont {Valenzuela}}, \bibinfo
  {author} {\bibfnamefont {E.}~\bibnamefont {D'Onghia}}, \ and\ \bibinfo
  {author} {\bibfnamefont {C.}~\bibnamefont {Firmani}},\ }\href {\doibase
  10.1086/322411} {\bibfield  {journal} {\bibinfo  {journal} {Astrophys. J.}\
  }\textbf {\bibinfo {volume} {559}},\ \bibinfo {pages} {516} (\bibinfo {year}
  {2001})},\ \Eprint {http://arxiv.org/abs/astro-ph/0010525}
  {arXiv:astro-ph/0010525} \BibitemShut {NoStop}%
\bibitem [{\citenamefont {Kamionkowski}\ and\ \citenamefont
  {Liddle}(2000)}]{Kamionkowski2000}%
  \BibitemOpen
  \bibfield  {author} {\bibinfo {author} {\bibfnamefont {M.}~\bibnamefont
  {Kamionkowski}}\ and\ \bibinfo {author} {\bibfnamefont {A.~R.}\ \bibnamefont
  {Liddle}},\ }\href {\doibase 10.1103/PhysRevLett.84.4525} {\bibfield
  {journal} {\bibinfo  {journal} {Phys. Rev. Lett.}\ }\textbf {\bibinfo
  {volume} {84}},\ \bibinfo {pages} {4525} (\bibinfo {year} {2000})},\ \Eprint
  {http://arxiv.org/abs/astro-ph/9911103} {arXiv:astro-ph/9911103} \BibitemShut
  {NoStop}%
\bibitem [{\citenamefont {Papastergis}\ \emph {et~al.}(2015)\citenamefont
  {Papastergis}, \citenamefont {Giovanelli}, \citenamefont {Haynes},\ and\
  \citenamefont {Shankar}}]{Papastergis2014}%
  \BibitemOpen
  \bibfield  {author} {\bibinfo {author} {\bibfnamefont {E.}~\bibnamefont
  {Papastergis}}, \bibinfo {author} {\bibfnamefont {R.}~\bibnamefont
  {Giovanelli}}, \bibinfo {author} {\bibfnamefont {M.~P.}\ \bibnamefont
  {Haynes}}, \ and\ \bibinfo {author} {\bibfnamefont {F.}~\bibnamefont
  {Shankar}},\ }\href {\doibase 10.1051/0004-6361/201424909} {\bibfield
  {journal} {\bibinfo  {journal} {Astron. Astrophys.}\ }\textbf {\bibinfo
  {volume} {574}},\ \bibinfo {pages} {A113} (\bibinfo {year} {2015})},\ \Eprint
  {http://arxiv.org/abs/1407.4665} {arXiv:1407.4665 [astro-ph.GA]} \BibitemShut
  {NoStop}%
\bibitem [{\citenamefont {Arvanitaki}\ \emph {et~al.}(2010)\citenamefont
  {Arvanitaki}, \citenamefont {Dimopoulos}, \citenamefont {Dubovsky},
  \citenamefont {Kaloper},\ and\ \citenamefont
  {March-Russell}}]{Arvanitaki2010}%
  \BibitemOpen
  \bibfield  {author} {\bibinfo {author} {\bibfnamefont {A.}~\bibnamefont
  {Arvanitaki}}, \bibinfo {author} {\bibfnamefont {S.}~\bibnamefont
  {Dimopoulos}}, \bibinfo {author} {\bibfnamefont {S.}~\bibnamefont
  {Dubovsky}}, \bibinfo {author} {\bibfnamefont {N.}~\bibnamefont {Kaloper}}, \
  and\ \bibinfo {author} {\bibfnamefont {J.}~\bibnamefont {March-Russell}},\
  }\href {\doibase 10.1103/PhysRevD.81.123530} {\bibfield  {journal} {\bibinfo
  {journal} {Phys. Rev. D}\ }\textbf {\bibinfo {volume} {81}},\ \bibinfo
  {pages} {123530} (\bibinfo {year} {2010})},\ \Eprint
  {http://arxiv.org/abs/0905.4720} {arXiv:0905.4720 [hep-th]} \BibitemShut
  {NoStop}%
\bibitem [{\citenamefont {Cicoli}(2013)}]{Cicoli2013}%
  \BibitemOpen
  \bibfield  {author} {\bibinfo {author} {\bibfnamefont {M.}~\bibnamefont
  {Cicoli}},\ }in\ \href {\doibase 10.3204/DESY-PROC-2013-04/cicoli_michele}
  {\emph {\bibinfo {booktitle} {{9th Patras Workshop on Axions, WIMPs and
  WISPs}}}}\ (\bibinfo {address} {Mainz, Germany},\ \bibinfo {year} {2013})\
  pp.\ \bibinfo {pages} {235--242},\ \Eprint {http://arxiv.org/abs/1309.6988}
  {arXiv:1309.6988 [hep-th]} \BibitemShut {NoStop}%
\bibitem [{\citenamefont {Peccei}\ and\ \citenamefont
  {Quinn}(1977)}]{Peccei1977}%
  \BibitemOpen
  \bibfield  {author} {\bibinfo {author} {\bibfnamefont {R.~D.}\ \bibnamefont
  {Peccei}}\ and\ \bibinfo {author} {\bibfnamefont {H.~R.}\ \bibnamefont
  {Quinn}},\ }\href {\doibase 10.1103/PhysRevLett.38.1440} {\bibfield
  {journal} {\bibinfo  {journal} {Phys. Rev. Lett.}\ }\textbf {\bibinfo
  {volume} {38}},\ \bibinfo {pages} {1440} (\bibinfo {year}
  {1977})}\BibitemShut {NoStop}%
\bibitem [{\citenamefont {Guth}\ \emph {et~al.}(2015)\citenamefont {Guth},
  \citenamefont {Hertzberg},\ and\ \citenamefont
  {Prescod-Weinstein}}]{Guth2015}%
  \BibitemOpen
  \bibfield  {author} {\bibinfo {author} {\bibfnamefont {A.~H.}\ \bibnamefont
  {Guth}}, \bibinfo {author} {\bibfnamefont {M.~P.}\ \bibnamefont {Hertzberg}},
  \ and\ \bibinfo {author} {\bibfnamefont {C.}~\bibnamefont
  {Prescod-Weinstein}},\ }\href {\doibase 10.1103/PhysRevD.92.103513}
  {\bibfield  {journal} {\bibinfo  {journal} {Phys. Rev. D}\ }\textbf {\bibinfo
  {volume} {92}},\ \bibinfo {pages} {103513} (\bibinfo {year} {2015})},\
  \Eprint {http://arxiv.org/abs/1412.5930} {arXiv:1412.5930 [astro-ph.CO]}
  \BibitemShut {NoStop}%
\bibitem [{\citenamefont {Dine}\ and\ \citenamefont
  {Fischler}(1983)}]{Dine1983}%
  \BibitemOpen
  \bibfield  {author} {\bibinfo {author} {\bibfnamefont {M.}~\bibnamefont
  {Dine}}\ and\ \bibinfo {author} {\bibfnamefont {W.}~\bibnamefont
  {Fischler}},\ }\href {\doibase 10.1016/0370-2693(83)90639-1} {\bibfield
  {journal} {\bibinfo  {journal} {Phys. Lett. B}\ }\textbf {\bibinfo {volume}
  {120}},\ \bibinfo {pages} {137} (\bibinfo {year} {1983})}\BibitemShut
  {NoStop}%
\bibitem [{\citenamefont {Preskill}\ \emph {et~al.}(1983)\citenamefont
  {Preskill}, \citenamefont {Wise},\ and\ \citenamefont
  {Wilczek}}]{Preskill1983}%
  \BibitemOpen
  \bibfield  {author} {\bibinfo {author} {\bibfnamefont {J.}~\bibnamefont
  {Preskill}}, \bibinfo {author} {\bibfnamefont {M.~B.}\ \bibnamefont {Wise}},
  \ and\ \bibinfo {author} {\bibfnamefont {F.}~\bibnamefont {Wilczek}},\ }\href
  {\doibase 10.1016/0370-2693(83)90637-8} {\bibfield  {journal} {\bibinfo
  {journal} {Phys. Lett. B}\ }\textbf {\bibinfo {volume} {120}},\ \bibinfo
  {pages} {127} (\bibinfo {year} {1983})}\BibitemShut {NoStop}%
\bibitem [{\citenamefont {Abbott}\ and\ \citenamefont
  {Sikivie}(1983)}]{Abbott1983}%
  \BibitemOpen
  \bibfield  {author} {\bibinfo {author} {\bibfnamefont {L.~F.}\ \bibnamefont
  {Abbott}}\ and\ \bibinfo {author} {\bibfnamefont {P.}~\bibnamefont
  {Sikivie}},\ }\href {\doibase 10.1016/0370-2693(83)90638-X} {\bibfield
  {journal} {\bibinfo  {journal} {Phys. Lett. B}\ }\textbf {\bibinfo {volume}
  {120}},\ \bibinfo {pages} {133} (\bibinfo {year} {1983})}\BibitemShut
  {NoStop}%
\bibitem [{\citenamefont {Kim}\ and\ \citenamefont {Carosi}(2010)}]{Kim2010}%
  \BibitemOpen
  \bibfield  {author} {\bibinfo {author} {\bibfnamefont {J.~E.}\ \bibnamefont
  {Kim}}\ and\ \bibinfo {author} {\bibfnamefont {G.}~\bibnamefont {Carosi}},\
  }\href {\doibase 10.1103/RevModPhys.82.557} {\bibfield  {journal} {\bibinfo
  {journal} {Rev. Mod. Phys.}\ }\textbf {\bibinfo {volume} {82}},\ \bibinfo
  {pages} {557} (\bibinfo {year} {2010})},\ \bibinfo {note} {[Erratum:
  Rev.Mod.Phys. 91, 049902 (2019)]},\ \Eprint {http://arxiv.org/abs/0807.3125}
  {arXiv:0807.3125 [hep-ph]} \BibitemShut {NoStop}%
\bibitem [{\citenamefont {Di~Luzio}\ \emph {et~al.}(2020)\citenamefont
  {Di~Luzio}, \citenamefont {Giannotti}, \citenamefont {Nardi},\ and\
  \citenamefont {Visinelli}}]{DiLuzio2020}%
  \BibitemOpen
  \bibfield  {author} {\bibinfo {author} {\bibfnamefont {L.}~\bibnamefont
  {Di~Luzio}}, \bibinfo {author} {\bibfnamefont {M.}~\bibnamefont {Giannotti}},
  \bibinfo {author} {\bibfnamefont {E.}~\bibnamefont {Nardi}}, \ and\ \bibinfo
  {author} {\bibfnamefont {L.}~\bibnamefont {Visinelli}},\ }\href {\doibase
  10.1016/j.physrep.2020.06.002} {\bibfield  {journal} {\bibinfo  {journal}
  {Phys. Rept.}\ }\textbf {\bibinfo {volume} {870}},\ \bibinfo {pages} {1}
  (\bibinfo {year} {2020})},\ \Eprint {http://arxiv.org/abs/2003.01100}
  {arXiv:2003.01100 [hep-ph]} \BibitemShut {NoStop}%
\bibitem [{\citenamefont {Alesini}\ \emph {et~al.}(2019)\citenamefont {Alesini}
  \emph {et~al.}}]{Alesini2019}%
  \BibitemOpen
  \bibfield  {author} {\bibinfo {author} {\bibfnamefont {D.}~\bibnamefont
  {Alesini}} \emph {et~al.},\ }\href {\doibase 10.1103/PhysRevD.99.101101}
  {\bibfield  {journal} {\bibinfo  {journal} {Phys. Rev. D}\ }\textbf {\bibinfo
  {volume} {99}},\ \bibinfo {pages} {101101} (\bibinfo {year} {2019})},\
  \Eprint {http://arxiv.org/abs/1903.06547} {arXiv:1903.06547
  [physics.ins-det]} \BibitemShut {NoStop}%
\bibitem [{\citenamefont {Rudelius}(2015)}]{Rudelius2015}%
  \BibitemOpen
  \bibfield  {author} {\bibinfo {author} {\bibfnamefont {T.}~\bibnamefont
  {Rudelius}},\ }\href {\doibase 10.1088/1475-7516/2015/9/020} {\bibfield
  {journal} {\bibinfo  {journal} {JCAP}\ }\textbf {\bibinfo {volume} {09}},\
  \bibinfo {pages} {020} (\bibinfo {year} {2015})},\ \Eprint
  {http://arxiv.org/abs/1503.00795} {arXiv:1503.00795 [hep-th]} \BibitemShut
  {NoStop}%
\bibitem [{\citenamefont {Hu}\ \emph {et~al.}(2000)\citenamefont {Hu},
  \citenamefont {Barkana},\ and\ \citenamefont {Gruzinov}}]{Hu2000}%
  \BibitemOpen
  \bibfield  {author} {\bibinfo {author} {\bibfnamefont {W.}~\bibnamefont
  {Hu}}, \bibinfo {author} {\bibfnamefont {R.}~\bibnamefont {Barkana}}, \ and\
  \bibinfo {author} {\bibfnamefont {A.}~\bibnamefont {Gruzinov}},\ }\href
  {\doibase 10.1103/PhysRevLett.85.1158} {\bibfield  {journal} {\bibinfo
  {journal} {Phys. Rev. Lett.}\ }\textbf {\bibinfo {volume} {85}},\ \bibinfo
  {pages} {1158} (\bibinfo {year} {2000})}\BibitemShut {NoStop}%
\bibitem [{\citenamefont {Hui}\ \emph {et~al.}(2017)\citenamefont {Hui},
  \citenamefont {Ostriker}, \citenamefont {Tremaine},\ and\ \citenamefont
  {Witten}}]{Hui2017}%
  \BibitemOpen
  \bibfield  {author} {\bibinfo {author} {\bibfnamefont {L.}~\bibnamefont
  {Hui}}, \bibinfo {author} {\bibfnamefont {J.~P.}\ \bibnamefont {Ostriker}},
  \bibinfo {author} {\bibfnamefont {S.}~\bibnamefont {Tremaine}}, \ and\
  \bibinfo {author} {\bibfnamefont {E.}~\bibnamefont {Witten}},\ }\href
  {\doibase 10.1103/PhysRevD.95.043541} {\bibfield  {journal} {\bibinfo
  {journal} {Phys. Rev. D}\ }\textbf {\bibinfo {volume} {95}},\ \bibinfo
  {pages} {043541} (\bibinfo {year} {2017})},\ \Eprint
  {http://arxiv.org/abs/1610.08297} {arXiv:1610.08297 [astro-ph.CO]}
  \BibitemShut {NoStop}%
\bibitem [{\citenamefont {Ringwald}(2012)}]{Ringwald2012}%
  \BibitemOpen
  \bibfield  {author} {\bibinfo {author} {\bibfnamefont {A.}~\bibnamefont
  {Ringwald}},\ }\href {\doibase 10.1016/j.dark.2012.10.008} {\bibfield
  {journal} {\bibinfo  {journal} {Phys. Dark Univ.}\ }\textbf {\bibinfo
  {volume} {1}},\ \bibinfo {pages} {116} (\bibinfo {year} {2012})},\ \Eprint
  {http://arxiv.org/abs/1210.5081} {arXiv:1210.5081 [hep-ph]} \BibitemShut
  {NoStop}%
\bibitem [{\citenamefont {{Madelung}}(1926)}]{Madelung1926}%
  \BibitemOpen
  \bibfield  {author} {\bibinfo {author} {\bibfnamefont {E.}~\bibnamefont
  {{Madelung}}},\ }\href {\doibase 10.1007/BF01504657} {\bibfield  {journal}
  {\bibinfo  {journal} {Naturwissenschaften}\ }\textbf {\bibinfo {volume}
  {14}},\ \bibinfo {pages} {1004} (\bibinfo {year} {1926})}\BibitemShut
  {NoStop}%
\bibitem [{\citenamefont {Schive}\ \emph
  {et~al.}(2014{\natexlab{a}})\citenamefont {Schive}, \citenamefont {Chiueh},\
  and\ \citenamefont {Broadhurst}}]{Schive2014}%
  \BibitemOpen
  \bibfield  {author} {\bibinfo {author} {\bibfnamefont {H.-Y.}\ \bibnamefont
  {Schive}}, \bibinfo {author} {\bibfnamefont {T.}~\bibnamefont {Chiueh}}, \
  and\ \bibinfo {author} {\bibfnamefont {T.}~\bibnamefont {Broadhurst}},\
  }\href {\doibase 10.1038/nphys2996} {\bibfield  {journal} {\bibinfo
  {journal} {Nature Phys.}\ }\textbf {\bibinfo {volume} {10}},\ \bibinfo
  {pages} {496} (\bibinfo {year} {2014}{\natexlab{a}})},\ \Eprint
  {http://arxiv.org/abs/1406.6586} {arXiv:1406.6586 [astro-ph.GA]} \BibitemShut
  {NoStop}%
\bibitem [{\citenamefont {Schive}\ \emph
  {et~al.}(2014{\natexlab{b}})\citenamefont {Schive}, \citenamefont {Liao},
  \citenamefont {Woo}, \citenamefont {Wong}, \citenamefont {Chiueh},
  \citenamefont {Broadhurst},\ and\ \citenamefont {Hwang}}]{Schive2014_2}%
  \BibitemOpen
  \bibfield  {author} {\bibinfo {author} {\bibfnamefont {H.-Y.}\ \bibnamefont
  {Schive}}, \bibinfo {author} {\bibfnamefont {M.-H.}\ \bibnamefont {Liao}},
  \bibinfo {author} {\bibfnamefont {T.-P.}\ \bibnamefont {Woo}}, \bibinfo
  {author} {\bibfnamefont {S.-K.}\ \bibnamefont {Wong}}, \bibinfo {author}
  {\bibfnamefont {T.}~\bibnamefont {Chiueh}}, \bibinfo {author} {\bibfnamefont
  {T.}~\bibnamefont {Broadhurst}}, \ and\ \bibinfo {author} {\bibfnamefont
  {W.~Y.~P.}\ \bibnamefont {Hwang}},\ }\href {\doibase
  10.1103/PhysRevLett.113.261302} {\bibfield  {journal} {\bibinfo  {journal}
  {Phys. Rev. Lett.}\ }\textbf {\bibinfo {volume} {113}},\ \bibinfo {pages}
  {261302} (\bibinfo {year} {2014}{\natexlab{b}})},\ \Eprint
  {http://arxiv.org/abs/1407.7762} {arXiv:1407.7762 [astro-ph.GA]} \BibitemShut
  {NoStop}%
\bibitem [{\citenamefont {Peebles}(2000)}]{Peebles2000}%
  \BibitemOpen
  \bibfield  {author} {\bibinfo {author} {\bibfnamefont {P.~J.~E.}\
  \bibnamefont {Peebles}},\ }\href {\doibase 10.1086/312677} {\bibfield
  {journal} {\bibinfo  {journal} {Astrophys. J. Lett.}\ }\textbf {\bibinfo
  {volume} {534}},\ \bibinfo {pages} {L127} (\bibinfo {year} {2000})},\ \Eprint
  {http://arxiv.org/abs/astro-ph/0002495} {arXiv:astro-ph/0002495} \BibitemShut
  {NoStop}%
\bibitem [{\citenamefont {Su\'arez}\ and\ \citenamefont
  {Chavanis}(2018)}]{Suarez2018}%
  \BibitemOpen
  \bibfield  {author} {\bibinfo {author} {\bibfnamefont {A.}~\bibnamefont
  {Su\'arez}}\ and\ \bibinfo {author} {\bibfnamefont {P.-H.}\ \bibnamefont
  {Chavanis}},\ }\href {\doibase 10.1103/PhysRevD.98.083529} {\bibfield
  {journal} {\bibinfo  {journal} {Phys. Rev. D}\ }\textbf {\bibinfo {volume}
  {98}},\ \bibinfo {pages} {083529} (\bibinfo {year} {2018})},\ \Eprint
  {http://arxiv.org/abs/1710.10486} {arXiv:1710.10486 [gr-qc]} \BibitemShut
  {NoStop}%
\bibitem [{\citenamefont {{Vaquero}}\ \emph {et~al.}(2019)\citenamefont
  {{Vaquero}}, \citenamefont {{Redondo}},\ and\ \citenamefont
  {{Stadler}}}]{2019JCAP...04..012V}%
  \BibitemOpen
  \bibfield  {author} {\bibinfo {author} {\bibfnamefont {A.}~\bibnamefont
  {{Vaquero}}}, \bibinfo {author} {\bibfnamefont {J.}~\bibnamefont
  {{Redondo}}}, \ and\ \bibinfo {author} {\bibfnamefont {J.}~\bibnamefont
  {{Stadler}}},\ }\href {\doibase 10.1088/1475-7516/2019/04/012} {\bibfield
  {journal} {\bibinfo  {journal} {JCAP}\ }\textbf {\bibinfo {volume} {2019}},\
  \bibinfo {eid} {012} (\bibinfo {year} {2019})},\ \Eprint
  {http://arxiv.org/abs/1809.09241} {arXiv:1809.09241 [astro-ph.CO]}
  \BibitemShut {NoStop}%
\bibitem [{\citenamefont {Kavanagh}\ \emph {et~al.}(2020)\citenamefont
  {Kavanagh}, \citenamefont {Edwards}, \citenamefont {Visinelli},\ and\
  \citenamefont {Weniger}}]{Kavanagh2020}%
  \BibitemOpen
  \bibfield  {author} {\bibinfo {author} {\bibfnamefont {B.~J.}\ \bibnamefont
  {Kavanagh}}, \bibinfo {author} {\bibfnamefont {T.~D.~P.}\ \bibnamefont
  {Edwards}}, \bibinfo {author} {\bibfnamefont {L.}~\bibnamefont {Visinelli}},
  \ and\ \bibinfo {author} {\bibfnamefont {C.}~\bibnamefont {Weniger}},\
  }\href@noop {} {\  (\bibinfo {year} {2020})},\ \Eprint
  {http://arxiv.org/abs/2011.05377} {arXiv:2011.05377 [astro-ph.GA]}
  \BibitemShut {NoStop}%
\bibitem [{\citenamefont {Chavanis}(2016)}]{Chavanis2016}%
  \BibitemOpen
  \bibfield  {author} {\bibinfo {author} {\bibfnamefont {P.-H.}\ \bibnamefont
  {Chavanis}},\ }\href {\doibase 10.1103/PhysRevD.94.083007} {\bibfield
  {journal} {\bibinfo  {journal} {Phys. Rev.}\ }\textbf {\bibinfo {volume}
  {D94}},\ \bibinfo {pages} {083007} (\bibinfo {year} {2016})},\ \Eprint
  {http://arxiv.org/abs/1604.05904} {arXiv:1604.05904 [astro-ph.CO]}
  \BibitemShut {NoStop}%
\bibitem [{\citenamefont {Fan}(2016)}]{Fan2016}%
  \BibitemOpen
  \bibfield  {author} {\bibinfo {author} {\bibfnamefont {J.}~\bibnamefont
  {Fan}},\ }\href {\doibase 10.1016/j.dark.2016.10.005} {\bibfield  {journal}
  {\bibinfo  {journal} {Phys. Dark Univ.}\ }\textbf {\bibinfo {volume} {14}},\
  \bibinfo {pages} {84} (\bibinfo {year} {2016})},\ \Eprint
  {http://arxiv.org/abs/1603.06580} {arXiv:1603.06580 [hep-ph]} \BibitemShut
  {NoStop}%
\bibitem [{\citenamefont {Edwards}\ \emph {et~al.}(2018)\citenamefont
  {Edwards}, \citenamefont {Kendall}, \citenamefont {Hotchkiss},\ and\
  \citenamefont {Easther}}]{Edwards2018}%
  \BibitemOpen
  \bibfield  {author} {\bibinfo {author} {\bibfnamefont {F.}~\bibnamefont
  {Edwards}}, \bibinfo {author} {\bibfnamefont {E.}~\bibnamefont {Kendall}},
  \bibinfo {author} {\bibfnamefont {S.}~\bibnamefont {Hotchkiss}}, \ and\
  \bibinfo {author} {\bibfnamefont {R.}~\bibnamefont {Easther}},\ }\href
  {\doibase 10.1088/1475-7516/2018/10/027} {\bibfield  {journal} {\bibinfo
  {journal} {JCAP}\ }\textbf {\bibinfo {volume} {10}},\ \bibinfo {pages} {027}
  (\bibinfo {year} {2018})},\ \Eprint {http://arxiv.org/abs/1807.04037}
  {arXiv:1807.04037 [astro-ph.CO]} \BibitemShut {NoStop}%
\bibitem [{\citenamefont {Eby}\ \emph {et~al.}(2020)\citenamefont {Eby},
  \citenamefont {Leembruggen}, \citenamefont {Street}, \citenamefont
  {Suranyi},\ and\ \citenamefont {Wijewardhana}}]{Eby2020}%
  \BibitemOpen
  \bibfield  {author} {\bibinfo {author} {\bibfnamefont {J.}~\bibnamefont
  {Eby}}, \bibinfo {author} {\bibfnamefont {M.}~\bibnamefont {Leembruggen}},
  \bibinfo {author} {\bibfnamefont {L.}~\bibnamefont {Street}}, \bibinfo
  {author} {\bibfnamefont {P.}~\bibnamefont {Suranyi}}, \ and\ \bibinfo
  {author} {\bibfnamefont {L.~C.~R.}\ \bibnamefont {Wijewardhana}},\ }\href
  {\doibase 10.1088/1475-7516/2020/10/020} {\bibfield  {journal} {\bibinfo
  {journal} {JCAP}\ }\textbf {\bibinfo {volume} {10}},\ \bibinfo {pages} {020}
  (\bibinfo {year} {2020})},\ \Eprint {http://arxiv.org/abs/2002.03022}
  {arXiv:2002.03022 [hep-ph]} \BibitemShut {NoStop}%
\bibitem [{\citenamefont {Oll\'e}\ \emph {et~al.}(2020)\citenamefont {Oll\'e},
  \citenamefont {Pujol\`as},\ and\ \citenamefont {Rompineve}}]{Olle2019}%
  \BibitemOpen
  \bibfield  {author} {\bibinfo {author} {\bibfnamefont {J.}~\bibnamefont
  {Oll\'e}}, \bibinfo {author} {\bibfnamefont {O.}~\bibnamefont {Pujol\`as}}, \
  and\ \bibinfo {author} {\bibfnamefont {F.}~\bibnamefont {Rompineve}},\ }\href
  {\doibase 10.1088/1475-7516/2020/02/006} {\bibfield  {journal} {\bibinfo
  {journal} {JCAP}\ }\textbf {\bibinfo {volume} {02}},\ \bibinfo {pages} {006}
  (\bibinfo {year} {2020})},\ \Eprint {http://arxiv.org/abs/1906.06352}
  {arXiv:1906.06352 [hep-ph]} \BibitemShut {NoStop}%
\bibitem [{\citenamefont {Kawasaki}\ \emph {et~al.}(2020)\citenamefont
  {Kawasaki}, \citenamefont {Nakano},\ and\ \citenamefont
  {Sonomoto}}]{Kawasaki2019}%
  \BibitemOpen
  \bibfield  {author} {\bibinfo {author} {\bibfnamefont {M.}~\bibnamefont
  {Kawasaki}}, \bibinfo {author} {\bibfnamefont {W.}~\bibnamefont {Nakano}}, \
  and\ \bibinfo {author} {\bibfnamefont {E.}~\bibnamefont {Sonomoto}},\ }\href
  {\doibase 10.1088/1475-7516/2020/01/047} {\bibfield  {journal} {\bibinfo
  {journal} {JCAP}\ }\textbf {\bibinfo {volume} {01}},\ \bibinfo {pages} {047}
  (\bibinfo {year} {2020})},\ \Eprint {http://arxiv.org/abs/1909.10805}
  {arXiv:1909.10805 [astro-ph.CO]} \BibitemShut {NoStop}%
\bibitem [{\citenamefont {Marsh}(2016)}]{Marsh2016}%
  \BibitemOpen
  \bibfield  {author} {\bibinfo {author} {\bibfnamefont {D.~J.~E.}\
  \bibnamefont {Marsh}},\ }\href {\doibase 10.1016/j.physrep.2016.06.005}
  {\bibfield  {journal} {\bibinfo  {journal} {Phys. Rept.}\ }\textbf {\bibinfo
  {volume} {643}},\ \bibinfo {pages} {1} (\bibinfo {year} {2016})},\ \Eprint
  {http://arxiv.org/abs/1510.07633} {arXiv:1510.07633 [astro-ph.CO]}
  \BibitemShut {NoStop}%
\bibitem [{\citenamefont {Baldeschi}\ \emph {et~al.}(1983)\citenamefont
  {Baldeschi}, \citenamefont {Ruffini},\ and\ \citenamefont
  {Gelmini}}]{Baldeschi1983}%
  \BibitemOpen
  \bibfield  {author} {\bibinfo {author} {\bibfnamefont {M.~R.}\ \bibnamefont
  {Baldeschi}}, \bibinfo {author} {\bibfnamefont {R.}~\bibnamefont {Ruffini}},
  \ and\ \bibinfo {author} {\bibfnamefont {G.~B.}\ \bibnamefont {Gelmini}},\
  }\href {\doibase 10.1016/0370-2693(83)90688-3} {\bibfield  {journal}
  {\bibinfo  {journal} {Phys. Lett. B}\ }\textbf {\bibinfo {volume} {122}},\
  \bibinfo {pages} {221} (\bibinfo {year} {1983})}\BibitemShut {NoStop}%
\bibitem [{\citenamefont {Ji}\ and\ \citenamefont {Sin}(1994)}]{Sin1994}%
  \BibitemOpen
  \bibfield  {author} {\bibinfo {author} {\bibfnamefont {S.~U.}\ \bibnamefont
  {Ji}}\ and\ \bibinfo {author} {\bibfnamefont {S.~J.}\ \bibnamefont {Sin}},\
  }\href {\doibase 10.1103/physrevd.50.3655} {\bibfield  {journal} {\bibinfo
  {journal} {Physical Review D}\ }\textbf {\bibinfo {volume} {50}},\ \bibinfo
  {pages} {3655–} (\bibinfo {year} {1994})}\BibitemShut {NoStop}%
\bibitem [{\citenamefont {Lee}\ and\ \citenamefont {Koh}(1996)}]{Lee1996}%
  \BibitemOpen
  \bibfield  {author} {\bibinfo {author} {\bibfnamefont {J.-w.}\ \bibnamefont
  {Lee}}\ and\ \bibinfo {author} {\bibfnamefont {I.-g.}\ \bibnamefont {Koh}},\
  }\href {\doibase 10.1103/PhysRevD.53.2236} {\bibfield  {journal} {\bibinfo
  {journal} {Phys. Rev. D}\ }\textbf {\bibinfo {volume} {53}},\ \bibinfo
  {pages} {2236} (\bibinfo {year} {1996})},\ \Eprint
  {http://arxiv.org/abs/hep-ph/9507385} {arXiv:hep-ph/9507385} \BibitemShut
  {NoStop}%
\bibitem [{\citenamefont {Matos}\ \emph {et~al.}(2000)\citenamefont {Matos},
  \citenamefont {Guzman},\ and\ \citenamefont {Urena-Lopez}}]{Matos2000}%
  \BibitemOpen
  \bibfield  {author} {\bibinfo {author} {\bibfnamefont {T.}~\bibnamefont
  {Matos}}, \bibinfo {author} {\bibfnamefont {F.~S.}\ \bibnamefont {Guzman}}, \
  and\ \bibinfo {author} {\bibfnamefont {L.~A.}\ \bibnamefont {Urena-Lopez}},\
  }\href {\doibase 10.1088/0264-9381/17/7/309} {\bibfield  {journal} {\bibinfo
  {journal} {Class. Quant. Grav.}\ }\textbf {\bibinfo {volume} {17}},\ \bibinfo
  {pages} {1707} (\bibinfo {year} {2000})},\ \Eprint
  {http://arxiv.org/abs/astro-ph/9908152} {arXiv:astro-ph/9908152} \BibitemShut
  {NoStop}%
\bibitem [{\citenamefont {Lee}(2016)}]{Lee2015}%
  \BibitemOpen
  \bibfield  {author} {\bibinfo {author} {\bibfnamefont {J.-W.}\ \bibnamefont
  {Lee}},\ }\href {\doibase 10.1016/j.physletb.2016.03.016} {\bibfield
  {journal} {\bibinfo  {journal} {Phys. Lett. B}\ }\textbf {\bibinfo {volume}
  {756}},\ \bibinfo {pages} {166} (\bibinfo {year} {2016})},\ \Eprint
  {http://arxiv.org/abs/1511.06611} {arXiv:1511.06611 [astro-ph.GA]}
  \BibitemShut {NoStop}%
\bibitem [{\citenamefont {Arbey}\ \emph {et~al.}(2001)\citenamefont {Arbey},
  \citenamefont {Lesgourgues},\ and\ \citenamefont {Salati}}]{Arbey2001}%
  \BibitemOpen
  \bibfield  {author} {\bibinfo {author} {\bibfnamefont {A.}~\bibnamefont
  {Arbey}}, \bibinfo {author} {\bibfnamefont {J.}~\bibnamefont {Lesgourgues}},
  \ and\ \bibinfo {author} {\bibfnamefont {P.}~\bibnamefont {Salati}},\ }\href
  {\doibase 10.1103/PhysRevD.64.123528} {\bibfield  {journal} {\bibinfo
  {journal} {Phys. Rev. D}\ }\textbf {\bibinfo {volume} {64}},\ \bibinfo
  {pages} {123528} (\bibinfo {year} {2001})},\ \Eprint
  {http://arxiv.org/abs/astro-ph/0105564} {arXiv:astro-ph/0105564} \BibitemShut
  {NoStop}%
\bibitem [{\citenamefont {Alcubierre}\ \emph {et~al.}(2002)\citenamefont
  {Alcubierre}, \citenamefont {Guzman}, \citenamefont {Matos}, \citenamefont
  {Nunez}, \citenamefont {Urena-Lopez},\ and\ \citenamefont
  {Wiederhold}}]{Alcubierre2001}%
  \BibitemOpen
  \bibfield  {author} {\bibinfo {author} {\bibfnamefont {M.}~\bibnamefont
  {Alcubierre}}, \bibinfo {author} {\bibfnamefont {F.~S.}\ \bibnamefont
  {Guzman}}, \bibinfo {author} {\bibfnamefont {T.}~\bibnamefont {Matos}},
  \bibinfo {author} {\bibfnamefont {D.}~\bibnamefont {Nunez}}, \bibinfo
  {author} {\bibfnamefont {L.~A.}\ \bibnamefont {Urena-Lopez}}, \ and\ \bibinfo
  {author} {\bibfnamefont {P.}~\bibnamefont {Wiederhold}},\ }\href {\doibase
  10.1088/0264-9381/19/19/314} {\bibfield  {journal} {\bibinfo  {journal}
  {Class. Quant. Grav.}\ }\textbf {\bibinfo {volume} {19}},\ \bibinfo {pages}
  {5017} (\bibinfo {year} {2002})},\ \Eprint
  {http://arxiv.org/abs/gr-qc/0110102} {arXiv:gr-qc/0110102} \BibitemShut
  {NoStop}%
\bibitem [{\citenamefont {Silverman}\ and\ \citenamefont
  {Mallett}(2002)}]{Silverman2002}%
  \BibitemOpen
  \bibfield  {author} {\bibinfo {author} {\bibfnamefont {M.~P.}\ \bibnamefont
  {Silverman}}\ and\ \bibinfo {author} {\bibfnamefont {R.~L.}\ \bibnamefont
  {Mallett}},\ }\href {\doibase 10.1023/A:1015934027224} {\bibfield  {journal}
  {\bibinfo  {journal} {Gen. Rel. Grav.}\ }\textbf {\bibinfo {volume} {34}},\
  \bibinfo {pages} {633} (\bibinfo {year} {2002})}\BibitemShut {NoStop}%
\bibitem [{\citenamefont {Arbey}\ \emph {et~al.}(2003)\citenamefont {Arbey},
  \citenamefont {Lesgourgues},\ and\ \citenamefont {Salati}}]{Arbey2003}%
  \BibitemOpen
  \bibfield  {author} {\bibinfo {author} {\bibfnamefont {A.}~\bibnamefont
  {Arbey}}, \bibinfo {author} {\bibfnamefont {J.}~\bibnamefont {Lesgourgues}},
  \ and\ \bibinfo {author} {\bibfnamefont {P.}~\bibnamefont {Salati}},\ }\href
  {\doibase 10.1103/PhysRevD.68.023511} {\bibfield  {journal} {\bibinfo
  {journal} {Phys. Rev. D}\ }\textbf {\bibinfo {volume} {68}},\ \bibinfo
  {pages} {023511} (\bibinfo {year} {2003})},\ \Eprint
  {http://arxiv.org/abs/astro-ph/0301533} {arXiv:astro-ph/0301533} \BibitemShut
  {NoStop}%
\bibitem [{\citenamefont {Boehmer}\ and\ \citenamefont
  {Harko}(2007)}]{Bohmer2007}%
  \BibitemOpen
  \bibfield  {author} {\bibinfo {author} {\bibfnamefont {C.~G.}\ \bibnamefont
  {Boehmer}}\ and\ \bibinfo {author} {\bibfnamefont {T.}~\bibnamefont
  {Harko}},\ }\href {\doibase 10.1088/1475-7516/2007/06/025} {\bibfield
  {journal} {\bibinfo  {journal} {JCAP}\ }\textbf {\bibinfo {volume} {06}},\
  \bibinfo {pages} {025} (\bibinfo {year} {2007})},\ \Eprint
  {http://arxiv.org/abs/0705.4158} {arXiv:0705.4158 [astro-ph]} \BibitemShut
  {NoStop}%
\bibitem [{\citenamefont {Fukuyama}\ \emph {et~al.}(2008)\citenamefont
  {Fukuyama}, \citenamefont {Morikawa},\ and\ \citenamefont
  {Tatekawa}}]{Fukuyama2008}%
  \BibitemOpen
  \bibfield  {author} {\bibinfo {author} {\bibfnamefont {T.}~\bibnamefont
  {Fukuyama}}, \bibinfo {author} {\bibfnamefont {M.}~\bibnamefont {Morikawa}},
  \ and\ \bibinfo {author} {\bibfnamefont {T.}~\bibnamefont {Tatekawa}},\
  }\href {\doibase 10.1088/1475-7516/2008/06/033} {\bibfield  {journal}
  {\bibinfo  {journal} {JCAP}\ }\textbf {\bibinfo {volume} {06}},\ \bibinfo
  {pages} {033} (\bibinfo {year} {2008})},\ \Eprint
  {http://arxiv.org/abs/0705.3091} {arXiv:0705.3091 [astro-ph]} \BibitemShut
  {NoStop}%
\bibitem [{\citenamefont {Sikivie}\ and\ \citenamefont
  {Yang}(2009)}]{Sikivie2009}%
  \BibitemOpen
  \bibfield  {author} {\bibinfo {author} {\bibfnamefont {P.}~\bibnamefont
  {Sikivie}}\ and\ \bibinfo {author} {\bibfnamefont {Q.}~\bibnamefont {Yang}},\
  }\href {\doibase 10.1103/PhysRevLett.103.111301} {\bibfield  {journal}
  {\bibinfo  {journal} {Phys. Rev. Lett.}\ }\textbf {\bibinfo {volume} {103}},\
  \bibinfo {pages} {111301} (\bibinfo {year} {2009})},\ \Eprint
  {http://arxiv.org/abs/0901.1106} {arXiv:0901.1106 [hep-ph]} \BibitemShut
  {NoStop}%
\bibitem [{\citenamefont {Lee}\ and\ \citenamefont {Lim}(2010)}]{Lee2010}%
  \BibitemOpen
  \bibfield  {author} {\bibinfo {author} {\bibfnamefont {J.-W.}\ \bibnamefont
  {Lee}}\ and\ \bibinfo {author} {\bibfnamefont {S.}~\bibnamefont {Lim}},\
  }\href {\doibase 10.1088/1475-7516/2010/01/007} {\bibfield  {journal}
  {\bibinfo  {journal} {JCAP}\ }\textbf {\bibinfo {volume} {01}},\ \bibinfo
  {pages} {007} (\bibinfo {year} {2010})},\ \Eprint
  {http://arxiv.org/abs/0812.1342} {arXiv:0812.1342 [astro-ph]} \BibitemShut
  {NoStop}%
\bibitem [{\citenamefont {Ruffini}\ and\ \citenamefont
  {Bonazzola}(1969)}]{Ruffini1969}%
  \BibitemOpen
  \bibfield  {author} {\bibinfo {author} {\bibfnamefont {R.}~\bibnamefont
  {Ruffini}}\ and\ \bibinfo {author} {\bibfnamefont {S.}~\bibnamefont
  {Bonazzola}},\ }\href {\doibase 10.1103/PhysRev.187.1767} {\bibfield
  {journal} {\bibinfo  {journal} {Phys. Rev.}\ }\textbf {\bibinfo {volume}
  {187}},\ \bibinfo {pages} {1767} (\bibinfo {year} {1969})}\BibitemShut
  {NoStop}%
\bibitem [{\citenamefont {Chavanis}\ and\ \citenamefont
  {Delfini}(2011)}]{Chavanis2011}%
  \BibitemOpen
  \bibfield  {author} {\bibinfo {author} {\bibfnamefont {P.~H.}\ \bibnamefont
  {Chavanis}}\ and\ \bibinfo {author} {\bibfnamefont {L.}~\bibnamefont
  {Delfini}},\ }\href {\doibase 10.1103/PhysRevD.84.043532} {\bibfield
  {journal} {\bibinfo  {journal} {Phys. Rev. D}\ }\textbf {\bibinfo {volume}
  {84}},\ \bibinfo {pages} {043532} (\bibinfo {year} {2011})},\ \Eprint
  {http://arxiv.org/abs/1103.2054} {arXiv:1103.2054 [astro-ph.CO]} \BibitemShut
  {NoStop}%
\bibitem [{\citenamefont {Amin}\ and\ \citenamefont {Mocz}(2019)}]{Amin2019}%
  \BibitemOpen
  \bibfield  {author} {\bibinfo {author} {\bibfnamefont {M.~A.}\ \bibnamefont
  {Amin}}\ and\ \bibinfo {author} {\bibfnamefont {P.}~\bibnamefont {Mocz}},\
  }\href {\doibase 10.1103/PhysRevD.100.063507} {\bibfield  {journal} {\bibinfo
   {journal} {Phys. Rev. D}\ }\textbf {\bibinfo {volume} {100}},\ \bibinfo
  {pages} {063507} (\bibinfo {year} {2019})},\ \Eprint
  {http://arxiv.org/abs/1902.07261} {arXiv:1902.07261 [astro-ph.CO]}
  \BibitemShut {NoStop}%
\bibitem [{\citenamefont {{Bao}}\ and\ \citenamefont {{Cai}}(2012)}]{Bao2013}%
  \BibitemOpen
  \bibfield  {author} {\bibinfo {author} {\bibfnamefont {W.}~\bibnamefont
  {{Bao}}}\ and\ \bibinfo {author} {\bibfnamefont {Y.}~\bibnamefont {{Cai}}},\
  }\href@noop {} {\bibfield  {journal} {\bibinfo  {journal} {arXiv e-prints}\
  ,\ \bibinfo {eid} {arXiv:1212.5341}} (\bibinfo {year} {2012})},\ \Eprint
  {http://arxiv.org/abs/1212.5341} {arXiv:1212.5341 [cond-mat.quant-gas]}
  \BibitemShut {NoStop}%
\bibitem [{\citenamefont {Desjacques}\ \emph {et~al.}(2018)\citenamefont
  {Desjacques}, \citenamefont {Kehagias},\ and\ \citenamefont
  {Riotto}}]{Desjacques2017}%
  \BibitemOpen
  \bibfield  {author} {\bibinfo {author} {\bibfnamefont {V.}~\bibnamefont
  {Desjacques}}, \bibinfo {author} {\bibfnamefont {A.}~\bibnamefont
  {Kehagias}}, \ and\ \bibinfo {author} {\bibfnamefont {A.}~\bibnamefont
  {Riotto}},\ }\href {\doibase 10.1103/PhysRevD.97.023529} {\bibfield
  {journal} {\bibinfo  {journal} {Phys. Rev. D}\ }\textbf {\bibinfo {volume}
  {97}},\ \bibinfo {pages} {023529} (\bibinfo {year} {2018})},\ \Eprint
  {http://arxiv.org/abs/1709.07946} {arXiv:1709.07946 [astro-ph.CO]}
  \BibitemShut {NoStop}%
\bibitem [{\citenamefont {Kirkpatrick}\ \emph {et~al.}(2020)\citenamefont
  {Kirkpatrick}, \citenamefont {Mirasola},\ and\ \citenamefont
  {Prescod-Weinstein}}]{Kirkpatrick2020}%
  \BibitemOpen
  \bibfield  {author} {\bibinfo {author} {\bibfnamefont {K.}~\bibnamefont
  {Kirkpatrick}}, \bibinfo {author} {\bibfnamefont {A.~E.}\ \bibnamefont
  {Mirasola}}, \ and\ \bibinfo {author} {\bibfnamefont {C.}~\bibnamefont
  {Prescod-Weinstein}},\ }\href {\doibase 10.1103/PhysRevD.102.103012}
  {\bibfield  {journal} {\bibinfo  {journal} {Phys. Rev. D}\ }\textbf {\bibinfo
  {volume} {102}},\ \bibinfo {pages} {103012} (\bibinfo {year} {2020})},\
  \Eprint {http://arxiv.org/abs/2007.07438} {arXiv:2007.07438 [hep-ph]}
  \BibitemShut {NoStop}%
\bibitem [{\citenamefont {Agrawal}(2019)}]{Agrawal2019}%
  \BibitemOpen
  \bibfield  {author} {\bibinfo {author} {\bibfnamefont {G.}~\bibnamefont
  {Agrawal}},\ }\href@noop {} {\emph {\bibinfo {title} {Nonlinear Fiber
  Optics}}}\ (\bibinfo  {publisher} {Academic Press},\ \bibinfo {address}
  {London, England},\ \bibinfo {year} {2019})\BibitemShut {NoStop}%
\bibitem [{\citenamefont {{Chavanis}}(2011)}]{Chavanis2011_2}%
  \BibitemOpen
  \bibfield  {author} {\bibinfo {author} {\bibfnamefont {P.-H.}\ \bibnamefont
  {{Chavanis}}},\ }\href {\doibase 10.1103/PhysRevD.84.043531} {\bibfield
  {journal} {\bibinfo  {journal} {\prd}\ }\textbf {\bibinfo {volume} {84}},\
  \bibinfo {eid} {043531} (\bibinfo {year} {2011})},\ \Eprint
  {http://arxiv.org/abs/1103.2050} {arXiv:1103.2050 [astro-ph.CO]} \BibitemShut
  {NoStop}%
\bibitem [{\citenamefont {Eby}\ \emph {et~al.}(2018)\citenamefont {Eby},
  \citenamefont {Leembruggen}, \citenamefont {Street}, \citenamefont
  {Suranyi},\ and\ \citenamefont {Wijewardhana}}]{Eby2018}%
  \BibitemOpen
  \bibfield  {author} {\bibinfo {author} {\bibfnamefont {J.}~\bibnamefont
  {Eby}}, \bibinfo {author} {\bibfnamefont {M.}~\bibnamefont {Leembruggen}},
  \bibinfo {author} {\bibfnamefont {L.}~\bibnamefont {Street}}, \bibinfo
  {author} {\bibfnamefont {P.}~\bibnamefont {Suranyi}}, \ and\ \bibinfo
  {author} {\bibfnamefont {L.~C.~R.}\ \bibnamefont {Wijewardhana}},\ }\href
  {\doibase 10.1103/PhysRevD.98.123013} {\bibfield  {journal} {\bibinfo
  {journal} {Phys. Rev. D}\ }\textbf {\bibinfo {volume} {98}},\ \bibinfo
  {pages} {123013} (\bibinfo {year} {2018})},\ \Eprint
  {http://arxiv.org/abs/1809.08598} {arXiv:1809.08598 [hep-ph]} \BibitemShut
  {NoStop}%
\bibitem [{\citenamefont {Eby}\ \emph {et~al.}(2016)\citenamefont {Eby},
  \citenamefont {Leembruggen}, \citenamefont {Suranyi},\ and\ \citenamefont
  {Wijewardhana}}]{Eby2016}%
  \BibitemOpen
  \bibfield  {author} {\bibinfo {author} {\bibfnamefont {J.}~\bibnamefont
  {Eby}}, \bibinfo {author} {\bibfnamefont {M.}~\bibnamefont {Leembruggen}},
  \bibinfo {author} {\bibfnamefont {P.}~\bibnamefont {Suranyi}}, \ and\
  \bibinfo {author} {\bibfnamefont {L.~C.~R.}\ \bibnamefont {Wijewardhana}},\
  }\href {\doibase 10.1007/JHEP12(2016)066} {\bibfield  {journal} {\bibinfo
  {journal} {JHEP}\ }\textbf {\bibinfo {volume} {12}},\ \bibinfo {pages} {066}
  (\bibinfo {year} {2016})},\ \Eprint {http://arxiv.org/abs/1608.06911}
  {arXiv:1608.06911 [astro-ph.CO]} \BibitemShut {NoStop}%
\bibitem [{\citenamefont {Chen}\ \emph {et~al.}(2020)\citenamefont {Chen},
  \citenamefont {Du}, \citenamefont {Lentz}, \citenamefont {Marsh},\ and\
  \citenamefont {Niemeyer}}]{Chen2020}%
  \BibitemOpen
  \bibfield  {author} {\bibinfo {author} {\bibfnamefont {J.}~\bibnamefont
  {Chen}}, \bibinfo {author} {\bibfnamefont {X.}~\bibnamefont {Du}}, \bibinfo
  {author} {\bibfnamefont {E.~W.}\ \bibnamefont {Lentz}}, \bibinfo {author}
  {\bibfnamefont {D.~J.~E.}\ \bibnamefont {Marsh}}, \ and\ \bibinfo {author}
  {\bibfnamefont {J.~C.}\ \bibnamefont {Niemeyer}},\ }\href@noop {} {\
  (\bibinfo {year} {2020})},\ \Eprint {http://arxiv.org/abs/2011.01333}
  {arXiv:2011.01333 [astro-ph.CO]} \BibitemShut {NoStop}%
\bibitem [{\citenamefont {Paredes}\ and\ \citenamefont
  {Michinel}(2016)}]{Paredes2015}%
  \BibitemOpen
  \bibfield  {author} {\bibinfo {author} {\bibfnamefont {A.}~\bibnamefont
  {Paredes}}\ and\ \bibinfo {author} {\bibfnamefont {H.}~\bibnamefont
  {Michinel}},\ }\href {\doibase 10.1016/j.dark.2016.02.003} {\bibfield
  {journal} {\bibinfo  {journal} {Phys. Dark Univ.}\ }\textbf {\bibinfo
  {volume} {12}},\ \bibinfo {pages} {50} (\bibinfo {year} {2016})},\ \Eprint
  {http://arxiv.org/abs/1512.05121} {arXiv:1512.05121 [astro-ph.CO]}
  \BibitemShut {NoStop}%
\bibitem [{\citenamefont {Rindler-Daller}\ and\ \citenamefont
  {Shapiro}(2012)}]{Rindler2011}%
  \BibitemOpen
  \bibfield  {author} {\bibinfo {author} {\bibfnamefont {T.}~\bibnamefont
  {Rindler-Daller}}\ and\ \bibinfo {author} {\bibfnamefont {P.~R.}\
  \bibnamefont {Shapiro}},\ }\href {\doibase 10.1111/j.1365-2966.2012.20588.x}
  {\bibfield  {journal} {\bibinfo  {journal} {Mon. Not. Roy. Astron. Soc.}\
  }\textbf {\bibinfo {volume} {422}},\ \bibinfo {pages} {135} (\bibinfo {year}
  {2012})},\ \Eprint {http://arxiv.org/abs/1106.1256} {arXiv:1106.1256
  [astro-ph.CO]} \BibitemShut {NoStop}%
\end{thebibliography}%

\end{document}